\documentclass[fleqn,usenatbib]{mnras}

\usepackage[T1]{fontenc}

\DeclareRobustCommand{\VAN}[3]{#2}
\let\VANthebibliography\thebibliography
\def\thebibliography{\DeclareRobustCommand{\VAN}[3]{##3}\VANthebibliography}

\usepackage{graphicx}	
\usepackage{amsmath}	
\usepackage{amssymb}	

\usepackage{newtxtext,newtxmath}

\title[HI observations in molecular clouds]{On the accuracy of  HI observations in molecular clouds -- More cold HI than thought?}
  \author[D. Seifried et al. ]
  {D.~Seifried,$^1$\thanks{seifried@ph1.uni-koeln.de} H.~Beuther,$^2$ S.~Walch,$^1$ J.~Syed,$^2$ J.~D.~Soler,$^2$
 P.~Girichidis,$^3$ R.~W\"unsch$^4$
 \\
  $^1$Universit\"at zu K\"oln, I. Physikalisches Institut, Z\"ulpicher Str. 77, 50937 K\"oln, Germany\\
  $^2$Max Planck Institute for Astronomy, K\"onigstuhl 17, 69117 Heidelberg, Germany \\
  $^3$Universit\"{a}t Heidelberg, Zentrum f\"{u}r Astronomie, Institut f\"{u}r Theoretische Astrophysik, Albert-Ueberle-Str. 2, 69120 Heidelberg, Germany\\
  $^4$Astronomical Institute of the Czech Academy of Sciences, Bocn\'{i} II 1401/1, 141 00 Prague 4, Czech Republic \\
  }

\date{Released 2021}

\pubyear{2021}

\begin{document}
\label{firstpage}
\pagerange{\pageref{firstpage}--\pageref{lastpage}}
\maketitle

\begin{abstract}
We present a study of the cold atomic hydrogen (HI) content of molecular clouds simulated within the SILCC-Zoom project for solar neighbourhood conditions. We produce synthetic observations of HI at 21~cm including HI self-absorption (HISA) and observational effects. We find that HI column densities, $N_\rmn{HI}$, of $\gtrsim$10$^{22}$~cm$^{-2}$ are frequently reached in molecular clouds with HI temperatures as low as $\sim$10~K. Hence, HISA observations assuming a fixed HI temperature tend to underestimate the amount of cold HI in molecular clouds by a factor of 3 -- 10 and produce an artificial upper limit of $N_\rmn{HI}$ around 10$^{21}$~cm$^{-2}$. We thus argue that the cold HI mass in molecular clouds could be a factor of a few higher than previously estimated. Also $N_\rmn{HI}$-PDFs obtained from HISA observations might be subject to observational biases and should be considered with caution. The underestimation of cold HI in HISA observations is due to both the large HI temperature variations and the effect of noise in regions of high optical depth. We find optical depths of cold HI around 1 -- 10 making optical depth corrections essential. We show that the high HI column densities ($\gtrsim$10$^{22}$~cm$^{-2}$) can in parts be attributed to the occurrence of up to 10 individual HI-H$_2$ transitions along the line of sight. This is also reflected in the spectra, necessitating Gaussian decomposition algorithms for their in-depth analysis. However, also for a single HI-H$_2$ transition, $N_\rmn{HI}$ frequently exceeds 10$^{21}$ cm$^{-2}$, challenging 1-dimensional, semi-analytical models. This is due to non-equilibrium chemistry effects and the fact that HI-H$_2$ transition regions usually do not possess a 1-dimensional geometry. Finally, we show that the HI gas is moderately supersonic with Mach numbers of a few. The corresponding non-thermal velocity dispersion can be determined via HISA observations within a factor of $\sim$2.
\end{abstract}




\section{Introduction}

The chemical composition of molecular clouds (MCs) is still a field of active research. In particular the transition from atomic hydrogen (HI) to molecular hydrogen (H$_2$) is of interest. During the initial formation of MCs, HI is continuously transformed into H$_2$ as the cloud collapses and its density increases \citep[e.g.][but see also the review by \citealt{Dobbs14}]{Glover07b,Glover10,Clark12b,MacLow12,Seifried17}. Knowledge about the exact content of HI and H$_2$ in MCs would also be of great value to assess the amount of CO-dark H$_2$ gas, i.e. molecular gas which is not traced by CO emission and thus affects the $X_\rmn{CO}$-factor \citep[see e.g. the review by][]{Bolatto13}.

Plenty of semi-analytical works studied the transition of HI- to H$_2$-dominated gas in MCs under various conditions with either plane-parallel or spherically symmetric models \citep[e.g.][]{Dishoeck86,Sternberg88,Roellig07,Krumholz08,Krumholz09,Wolfire10,Sternberg14,Bialy16}. For solar-neighbourhood conditions these models predict that the HI-H$_2$ transition occurs around column densities of \mbox{$\sim$10$^{20}$ -- 10$^{21}$ cm$^{-2}$}. Furthermore, beside the column density where the \textit{transition} occurs, \citet{Sternberg14} and \cite{Bialy16} also find that the \textit{total} HI column density, $N_\rmn{HI}$, should be limited to a maximum of about 10$^{21}$~cm$^{-2}$. These models are, however, typically highly idealised as they assume chemical equilibrium and are limited to a 1D geometry, although attempts are made to extend them to turbulent environments \citep{Bialy17b}.

Modelling of the HI-H$_2$ transition in 3D, magneto-hydrodynamical (MHD) simulations remains a highly challenging task. This is due to that fact that MCs are not necessarily in chemical equilibrium concerning their hydrogen content due to turbulent mixing of H$_2$ into the lower-density environment \citep{Glover10,Valdivia16,Seifried17}. Therefore, any simulation trying to study the physics and chemistry of the HI-H$_2$ transition in a self-consistent manner requires the inclusion of an on-the-fly chemical network. Moreover, in order to achieve a converged H$_2$ (and thus HI) content in the simulations, a high spatial resolution of $\sim$0.1~pc is required \citep{Seifried17}, which was confirmed subsequently by means of semi-analytical considerations \citep{Joshi19}. Despite these difficulties, there have been a great number of studies attempting to model the chemical composition of the (dense) interstellar medium (ISM) \citep[e.g.][and many more]{Gnedin09,Glover10,MacLow12,Valdivia16,Bialy17b,Clark19,Joshi19,Nickerson19,Bellomi20,Smith20}. However, not all of these studies fulfill the requirements of non-equilibrium chemistry and a sufficient spatial resolution. In agreement with semi-analytical results the HI-H$_2$ transition is found to occur around \mbox{$\sim$10$^{20}$ -- 10$^{21}$ cm$^{-2}$} \citep{Gnedin09,Valdivia16,Seifried17,Bellomi20}.

From the observational perspective, the HI content in the ISM is often determined via observations of the HI~21~cm emission line, UV absorption measurements and far-infrared studies. A large number of observations on both Galactic \citep[e.g.][]{Savage77,Kalberla05,Gillmon06,Rachford09,Barriault10,Stanimirovic14,Lee12,Lee15,Burkhart15,Imara16} and extragalactic scales \citep[e.g.][]{Wong02,Browning03,Blitz04,Blitz06,Bigiel08,Wong09,Schruba11} find that the HI-H$_2$ transition occurs around \mbox{$\sim$10$^{20}$ -- 10$^{21}$~cm$^{-2}$}. Furthermore, beside the value for the \textit{transition}, some of these observations \citep[e.g.][]{Wong02,Bigiel08,Barriault10,Schruba11,Lee12,Stanimirovic14,Burkhart15} also suggest an upper \textit{threshold} of $N_\rmn{HI}$ around 10$^{21}$~cm$^{-2}$, similar to the aforementioned semi-analytical models.

On smaller scales of individual MCs, the measurement of their cold HI content via the HI 21~cm line is challenging due to the simultaneous emission of HI in the warm neutral medium. This problem can be overcome by the study of HI self-absorption (HISA) first reported by \citet{Heeschen54,Heeschen55}. These HISA features arise when cold HI in the foreground absorbs the emission of warmer HI in the background \citep[e.g.][]{Knapp74}. Over the last decades there have been numerous HISA observations studying the properties of HI gas in MCs \citep[e.g.][]{Riegel72,Knapp74,vanderWerf88,Goodman94,Montgomery95,Gibson00,Gibson05,Kavars03,Kavars05,Li03,Goldsmith05,Klaassen05,Krco08,Krco10,Denes18,Beuther20,Syed20,Wang20}. Similar to the observations on galactic scales and semi-analytical models, a large number of these HISA observations towards MCs seem to confirm an upper column density threshold of a few 10$^{21}$~cm$^{-2}$ for cold HI. However, there have been observations which report partly significantly higher HI column densities up to $\sim$10$^{22}$~cm$^{-2}$ \citep{Motte14,Bihr15,Denes18}. Also some indirect measurements suggest higher HI column densities \citep{Fukui14,Fukui15,Okamoto17}. These observations thus challenge the picture of a saturation of $N_\rmn{HI}$ around 10$^{21}$~cm$^{-2}$ obtained from other observational works and semi-analytical methods.

Some of the first synthetic HI observations are presented by \citet{Douglas10} and \citet{Acreman10,Acreman12}, modelling the HI emission on \textit{galactic} scales with resolutions around a few 1~pc, i.e. stemming from the more diffuse, atomic ISM. On similar scales, \citet{Kim14}, and in associated follow-up studies \citet{Murray15,Murray17}, conclude that HI absorption observations can trace parameters like the $N_\rmn{HI}$ and the spin temperature with an accuracy of a few 10\%. However, the authors do not include a chemical network for HI-H$_2$ and only apply a simplified radiative transfer method without observational effects such as noise or beam smearing. On smaller scales of individual MCs, both \citet{Hennebelle07} and \citet{Heiner15} present HI spectra from simulations without any chemical network and only a simplified radiative transfer method. However, on these scales HI and H$_2$ coexist and are out of equilibrium due to turbulent mixing \citep{Glover07b,Valdivia16,Seifried17} and thus the predictive power of these studies is limited. \citet{Fukui18} present synthetic HI spectra of colliding flow simulations including a chemical network for HI and H$_2$. The authors find that for a significant portion of the pixels, $N_\rmn{HI}$ can be underestimated in a non-negligible manner by several 10\% and more, in particular in pixels with high optical depths. They do, however, not report any HI self-absorption features nor include observational effects. Furthermore, as the authors consider a MC at an extremely early evolutionary stage, their findings rather apply to the cold and warm neutral medium. Finally, \citet{Soler19} apply the method of oriented gradients to synthetic HISA observations of MCs modelled by \citet{Clark19}.

Following these works, here we present fully self-consistent synthetic HISA observations of MCs, that is, including 3D, MHD simulations with a chemical network, high spatial resolution ($\sim$0.1~pc), self-consistent radiative calculations and observational effects. We investigate the accuracy of HISA observations towards MCs (Section~\ref{sec:results}) and discuss implications of our findings for the postulated saturation of $N_\rmn{HI}$ around 10$^{21}$~cm$^{-2}$ (Section~\ref{sec:discussion}). In particular we show that this saturation of $N_\rmn{HI}$ could be a purely observational effect and/or a consequence of oversimplifying assumptions in semi-analytical models (Section~\ref{sec:moreHI}). We conclude our work in Section~\ref{sec:conclusion}.

\section{Numerics}

\subsection{Simulations}

We here briefly describe the numerics behind the simulations, for more details we refer to \citet{Seifried17,Seifried19}. All simulations are performed with the FLASH code \citep{Fryxell00,Dubey08}. We include a chemical network for H$_2$, H, H$^+$, C$^+$, C, O, CH, OH, CO, HCO$^+$, He, He$^+$, M and M$^+$ \citep{Nelson99,Glover07a,Glover07b,Glover12} with updates in the network described in \citet{Mackey19}. Here, M and M$^+$ represent the contribution of metals, where we specifically consider Si and its first ionised state. We also include the most relevant heating and cooling mechanisms. In addition, we calculate the attenuation of the interstellar radiation field \citep[$G_0$ = 1.7 in Habing units,][]{Habing68,Draine78} using the \textsc{TreeRay/OpticalDepth} module \citep{Clark12,Walch15,Wunsch18}. The cosmic ray ionisation rate of atomic hydrogen is set to \mbox{3 $\times$ 10$^{-17}$ s$^{-1}$}. The Poisson equation for self-gravity is solved using a tree-based method \citep{Wunsch18}.

The simulations are part of the SILCC-Zoom project \citep{Seifried17,Seifried20}, where we model the formation and evolution of MCs located in a part of a stratified galactic disk, which in turn is part of the SILCC project \citep{Walch15,Girichidis16}. The disk has an initial Gaussian density profile given by
\begin{equation}
 \rho(z) = \rho_0 \times \textrm{exp}\left[ - \frac{1}{2} \left( \frac{z}{h_z} \right)^2 \right] \, ,
 \label{eq:rhosilcc}
\end{equation}
with $h_z$ = 30 pc and $\rho_0 = 9 \times 10^{-24}$ g cm$^{-3}$, resulting in a total gas surface density of \mbox{$\Sigma_\rmn{gas}$ = 10 M$_{\sun}$ pc$^{-2}$}. We run two simulations, one without a magnetic field and one with. For the magnetized run we initialise the magnetic field as 
\begin{equation}
 B_{x} = B_{x,0} \sqrt{\rho(z)/\rho_0} \; , B_y = 0 \; , B_z = 0 \, ,
 \label{eq:bsilcc}
\end{equation}
with $B_{x,0}$ = 3 $\mu$G in accordance with observations \citep{Beck13}. We emphasise that the magnetic field is dynamically important for the (chemical) evolution of the MCs \citep{Seifried20,Seifried20b}. In addition to the gas self-gravity we include a background potential from the old stellar component modelled as an isothermal sheet with a scale height of 100~pc and \mbox{$\Sigma_\rmn{star}$ = 30 M$_{\sun}$ pc$^{-2}$}.

In the initial simulation phase, up to a time $t_0$, the spatial resolution is 4~pc and we drive turbulence by injecting supernovae (SNe) with a rate of 15~SNe~Myr$^{-1}$ \citep[see][for details]{Walch15,Girichidis16,Gatto17}. At $t_0$ we stop the SN injection to allow for the formation of MCs unaffected by nearby SN remnants, which could influence their evolution \citep{Seifried18}. For both the unmagnetized and the magnetized run, we pick two regions, each with a typical size of $\sim$(100~pc)$^3$, in which MCs are about to form. We thus have in total 4 MCs, henceforth denoted as MC1-HD, MC2-HD, MC1-MHD and MC2-MHD, where the first two are non-magnetised runs and the latter two include a dynamically relevant magnetic field\footnote{See also \citet{Seifried20}, where we discuss their chemical properties concerning H$_2$ and CO. In that publication, the MHD clouds were named MC3-MHD and MC4-MHD. Also note that, due to the nature of the SILCC simulations, the runs including magnetic fields are completely independent from those without magnetic fields, and so are the resulting clouds.} The typical H$_2$ masses of these clouds are around \mbox{20 - 50 $\times$ $10^3$ M$_{\sun}$}, and for the magnetized runs the volume-weighted magnetic field is around 4~$\mu$G. Starting at $t_0$, we then progressively increase the spatial resolution in these zoom-in regions over 1.65~Myr reaching a maximum resolution of 0.06~pc. Afterwards, we evolve the clouds on this resolution for a few more Myr. We note that throughout the paper all times refer to the time elapsed since $t_0$, i.e. the start of the zoom-in procedure. We have \mbox{$t_0$ = 11.9 and 16.0 Myr} for the runs without and with magnetic fields, respectively.

\subsection{Radiative transfer}
\label{sec:radtrans}

The radiative transfer simulations are performed in a post-processing step with RADMC-3D \citep{Dullemond12} for the H\ion{I}~21~cm emission line of atomic hydrogen. In order to calculate the (two-) level population, we apply the method to calculate the spin temperature of atomic hydrogen, $T_\rmn{s}$, described in \citet[][their Eqs. 4 -- 7]{Kim14}. In particular, we include the Wouthuysen-Field (WF) effect \citep{Wouthuysen52,Field58} assuming \mbox{$T_{\alpha}$ = $T_\rmn{gas}$} for the effective temperature of the Ly$\alpha$ field \citep{Field59} and \mbox{$n_{\alpha}$ = 10$^6$ cm$^{-3}$} for the Ly$\alpha$ photon density \citep{Liszt01}. We emphasise that including or excluding the WF effect has only a marginal impact as for the temperatures of $\lesssim$10$^2$~K typical for MCs, the WF effect does barely affect $T_\rmn{s}$, which remains close to the actual gas temperature in both cases \citep[fig. 2 of][]{Kim14}. Furthermore, the Einstein coefficient of the HI line is \mbox{$A_{ul}$ = 2.8843 $\times$ 10$^{-15}$ s$^{-1}$} \citep{Gould94} and the spectral resolution is set to \mbox{200 m s$^{-1}$} over a range of \mbox{$\pm$20 km s$^{-1}$} resulting in 201 velocity channels.

We consider the emission of the four MCs in isolation, i.e. the emission stemming from the gas in the aforementioned zoom-in regions only. This allows us to focus on the HISA signal originating from the MCs themselves (and to a smaller extent from warm HI in the zoom-in region) avoiding any foreground contamination. We use a resolution of 0.06~pc, i.e. identical to the maximum resolution of the underlying simulations. We investigate the emission for two points in time, that is \mbox{$t_\rmn{evol}$ = 2} and 3~Myr. As the results for both times are qualitatively similar, for most of the plots we focus on \mbox{$t_\rmn{evol}$ = 2 Myr}. In order to model the emission (and its absorption) of a diffuse HI background, we include a (spatially and spectrally) fixed background radiation field with a brightness temperature of 100~K in the radiative transfer calculation. This background temperature is motivated by results from recent HISA observations within the galactic plane \citep{Syed20,Wang20} and is also used in the numerical study presented in \citet{Soler19b}. This makes our HISA observations sensitive to absorption of HI gas with $T_\rmn{s}$ below 100~K, warmer gas will be seen in emission. We note that we have also tested the usage of a background temperature of 200~K. The observed changes are, however, only very moderate which is why we focus on the case of 100~K here.

\subsubsection{Adding observational effects}
\label{sec:obs_effects}

In a final step we incorporate observational effects into our obtained (ideal) HI emission maps. For this purpose we (i) convolve our emission maps with a Gaussian beam of 80'' (at the chosen distance, see below), (ii) reduce the spectral resolution from \mbox{200 m s$^{-1}$} to \mbox{1 km s$^{-1}$} by summing up the contribution of 5~neighbouring channels, i.e. the data cubes analysed in the following have only 40 velocity channels (one fifth of the original cubes), and (iii) finally add random Gaussian noise with a standard deviation of 3~K to the obtained emission maps\footnote{Note that it is important to apply step (iii) as the final step.}. All the above stated values are average values from recent HISA observations towards MCs \citep[e.g.][]{Denes18,Beuther20,Syed20,Wang20}. We choose two different distances for the observed clouds of 150~pc and 3~kpc, corresponding to a physical beam size of 0.06~pc and 1.2~pc, respectively. However, as the results for both distances are relatively similar, in the following we focus on the distance of 150~pc.

\subsection{HISA calculations}
\label{sec:HISA_calc}

In order to investigate the properties of the absorbing atomic hydrogen gas, i.e. the HISA features, we follow the approach described in \citet[][see their section 2.2]{Wang20} to convert the absorption spectrum into an effective emission spectrum of the cold HI. In short, assuming a temperature of the absorbing, cold HI gas, $T_\rmn{HISA}$, and that both foreground and background emission are optically thin, one can relate the observed emission to the optical depth, $\tau_\rmn{HISA}$, of the absorption layer via 
\begin{equation}
  T_\rmn{off-on} = T_\rmn{off} - T_\rmn{on} = (p T_\rmn{off} \, + \, T_\rmn{cont} \, - \,  T_\rmn{HISA} )\times(1 - \rmn{e}^{-\tau_\rmn{HISA}}) \, .
 \label{eq:Tonoff}
\end{equation}
For the sake of readability, we have omitted the dependence on the velocity channel in the above equation. Here, $T_\rmn{on}$ is the actually observed brightness temperature, and $T_\rmn{off}$ is the brightness temperature at an off-position, i.e. the brightness temperature which would be measured if no absorbing cold HI gas were present along the line of sight (LOS) towards the observer. Furthermore, $T_\rmn{cont}$ is the brightness temperature of the diffuse continuum background, and the dimensionless quantity $p$ parametrises the ratio of foreground to background emission \citep{Feldt93,Gibson00} and is usually estimated to be close to 1 \citep{McClure06,Rebolledo17,Denes18,Wang20}.

In \citet{Wang20}, $T_\rmn{off}$ -- which is not accessible via observations -- is inferred by fitting a polynomial to the absorption-free channels, i.e. those channels where no HISA feature is present. As we use a fixed background brightness temperature of 100~K in our synthetic observations, which varies neither spatially nor spectrally. Therefore, we cannot test additional inaccuracies arising from the uncertainty to determine $T_\rmn{off}$ in actual observations -- a caveat to keep in mind throughout the paper. However, due to this simplifying assumption, in our case the spectrum in the absorption-free channels is practically flat as the emission of HI gas warmer than 100~K in the considered zoom-in region is almost negligible, similar to the findings of \citet{Soler19b}. For this reason, we can set \mbox{$T_\rmn{off}$ = 100 K} for our data which yields
\begin{equation}
 T_\rmn{off-on}(v) = 100 \, \rmn{K} \,- \, T_\rmn{on}(v) \, .
 \label{eq:Tonoff_inv}
\end{equation}
The quantity $T_\rmn{off-on}(v)$ is positive and gives the depth of the absorption feature seen in $T_\rmn{on}(v)$. In the following we refer to these kind of spectra as HISA spectra.

Furthermore, using a constant background temperature and assuming that the emission of HI warmer than 100~K is negligible is congruent with setting $p$ = 1 and \mbox{$p$$T_\rmn{off}$ + $T_\rmn{cont}$ = 100~K}\footnote{Effectively, we now do not differentiate any more between $p T_\rmn{off}$ and $T_\rmn{cont}$.}. Hence, we can simplify Eq.~\ref{eq:Tonoff} to yield
\begin{equation}
 T_\rmn{off-on}(v) = (100 \, \rmn{K} - T_\rmn{HISA}) \times (1 - \rmn{e}^{-\tau_\rmn{HISA}(v)}) \, .
 \label{eq:Toffon}
\end{equation}

We apply the spectral analysis tool \textsc{BTS}\footnote{Acronym for ``Behind The Spectrum'', \url{https://github.com/SeamusClarke/BTS}} \citep{Clarke18} to $T_\rmn{off-on}(v)$ to find the location and properties of the HISA feature. BTS identifies and fits Gaussian peaks in a spectrum. Assuming that the HISA feature has approximately the shape of a single Gaussian, we restrict the fitting function used in BTS to a single Gaussian. We set the noise level to 3~K and the required signal-to-noise ratio to 3. Of the obtained fitting values, here only the line width, \mbox{$\sigma_\rmn{BTS}$}, will be used later in the paper. Next, for each pixel for which \textsc{BTS} identifies a (Gaussian) HISA feature, we calculate the column density of the cold HI responsible for the HISA feature.

For this purpose, for most of our paper we adopt a fixed value for $T_\rmn{HISA}$ for the entire map, identical to the approach applied in recent observations \citep[e.g.][but see below and Section~\ref{sec:T_HISA_free} for a different approach]{Syed20,Wang20}. We then solve Eq.~\ref{eq:Toffon} for the optical depth of the HISA feature, $\tau_\rmn{HISA}$, for each channel independently, which yields
\begin{equation}
 \tau_\rmn{HISA}(v) = -\textrm{ln} \left( 1 - \frac{T_\rmn{off-on}}{100 \, \rmn{K}  - T_\rmn{HISA}} \right) \, .
\label{eq:tau_HISA}
\end{equation}
With this we can calculate the column density of the HISA feature via \citep{Wilson13}
\begin{equation}
 N_\rmn{HI, obs} = 1.8224 \times 10^{18} \text{cm$^{-2}$} \, \frac{T_\rmn{s}}{\text{1 K}} \int \tau(v) \frac{\rmn{d}v} { \textrm{1 km s$^{-1}$} } \, ,
\label{eq:NH}
\end{equation}
where we assume \mbox{$T_\rmn{s}$ = $T_\rmn{HISA}$} for the HI spin temperature and \mbox{$\tau(v)$ = $\tau_\rmn{HISA}(v)$} for the optical depth.

Though being straightforward to use, this method has the disadvantage that a fixed value of $T_\rmn{HISA}$ for every pixel of the map has to be assumed, which might not be the case. Furthermore, when choosing a fixed $T_\rmn{HISA}$, Eq.~\ref{eq:Tonoff} might not yield a result for $\tau_\rmn{HISA}$ for every channel. The maximum useable value for each channel, $T_\rmn{HISA, max}(v)$, is obtained by assuming \mbox{$\tau_\rmn{HISA}$ $\rightarrow$ $\infty$} \citep{Wang20}, which results for our setup in
\begin{equation}
T_\rmn{HISA, max}(v) = -T_\rmn{off-on}(v) + p T_\rmn{off}(v) + T_\rmn{cont} = T_\rmn{on}(v) \, .
\label{eq:Tmax}
\end{equation}
Hence, if the assumed $T_\rmn{HISA}$ exceeds the observed brightness temperature $T_\rmn{on}(v)$ of a given channel, this channel has to be dropped and cannot be taken into account for the calculation of the column density (see Section~\ref{sec:HI_temp}).

An alternative way to determine the optical depth, which simultaneously leaves $T_\rmn{HISA}$ as a free parameter, is given by \citet[][]{Knapp74} expressing $\tau_\rmn{HISA}(v)$ via
\begin{equation}
\tau_\rmn{HISA}(v) = \tau_0 e^{-\frac{1}{2} \left( \frac{v-v_0}{\sigma_0} \right)^2} \, .
\label{eq:tau_Knapp}
\end{equation}
Here, $\tau_0$, $v_0$ and $\sigma_0$ are free parameters (together with $T_\rmn{HISA}$) which are determined by fitting the observed spectrum $T_\rmn{off-on}(v)$ with Eq.~\ref{eq:tau_Knapp} inserted in Eq.~\ref{eq:Toffon}. As before, the fit is only performed for those pixels where we identify a Gaussian HISA feature with BTS. Beside the approach using a fixed $T_\rmn{HISA}$ we will also test this approach in the following.

\section{Results}
\label{sec:results}

\begin{figure}
\centering
\includegraphics[width=\linewidth]{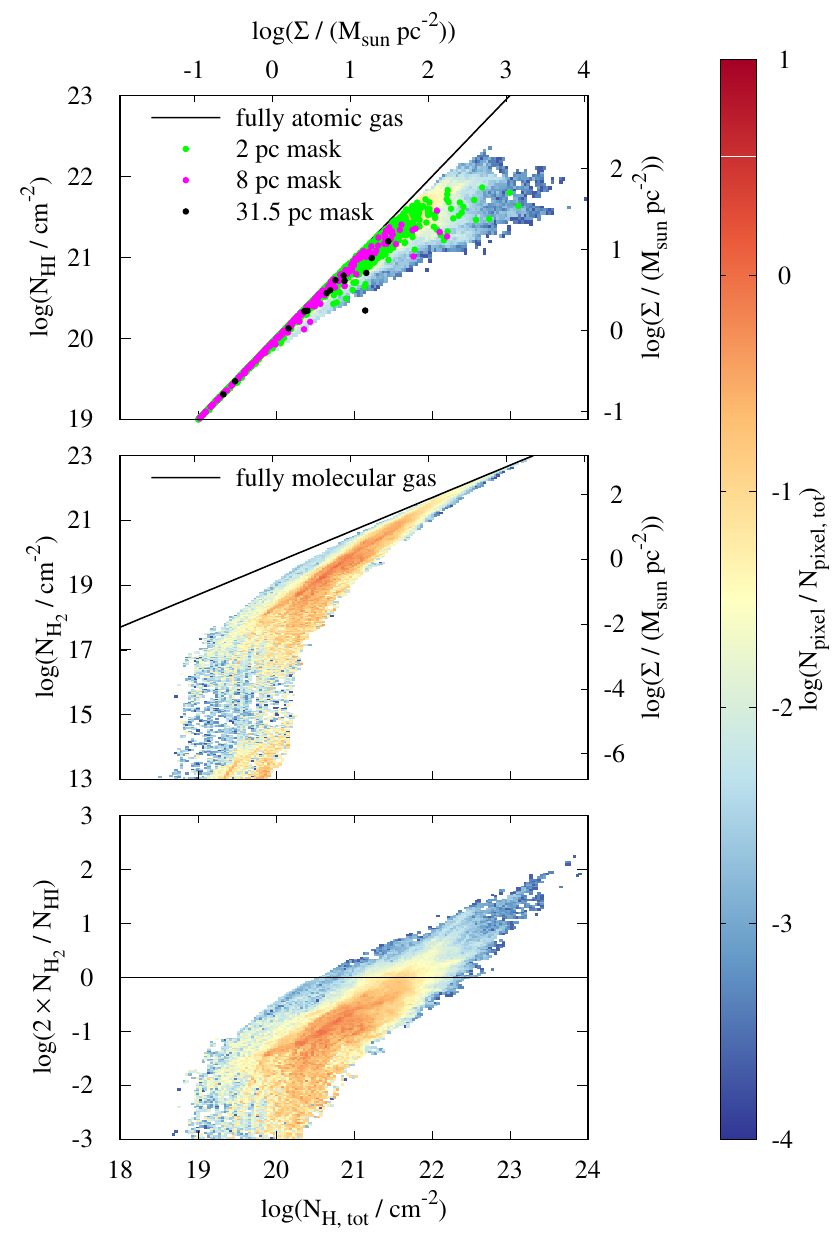} 
\caption{2D-PDF of $N_\rmn{HI}$ (top) and $N_\rmn{H_2}$ (middle) and the ratio of both column densities vs. $N_\rmn{H, tot}$ (bottom) for MC1-HD at 3~Myr for one LOS. The HI-H$_2$ transition occurs around 10$^{21}$~cm$^{-2}$ in rough agreement with observational results. However, the HI column density levels off only around a few 10$^{22}$~cm$^{-2}$, thus higher than typically obtained in observations. Colored dots denote the average column density when using pixels with a side length of 2, 8 and 31.5~pc (see Section~\ref{sec:moreHI}). Note the different $y$-axis scaling for the top and middle panel.}
\label{fig:NH-diag}
\end{figure}

\begin{figure*}
\centering
\includegraphics[width=0.9\linewidth]{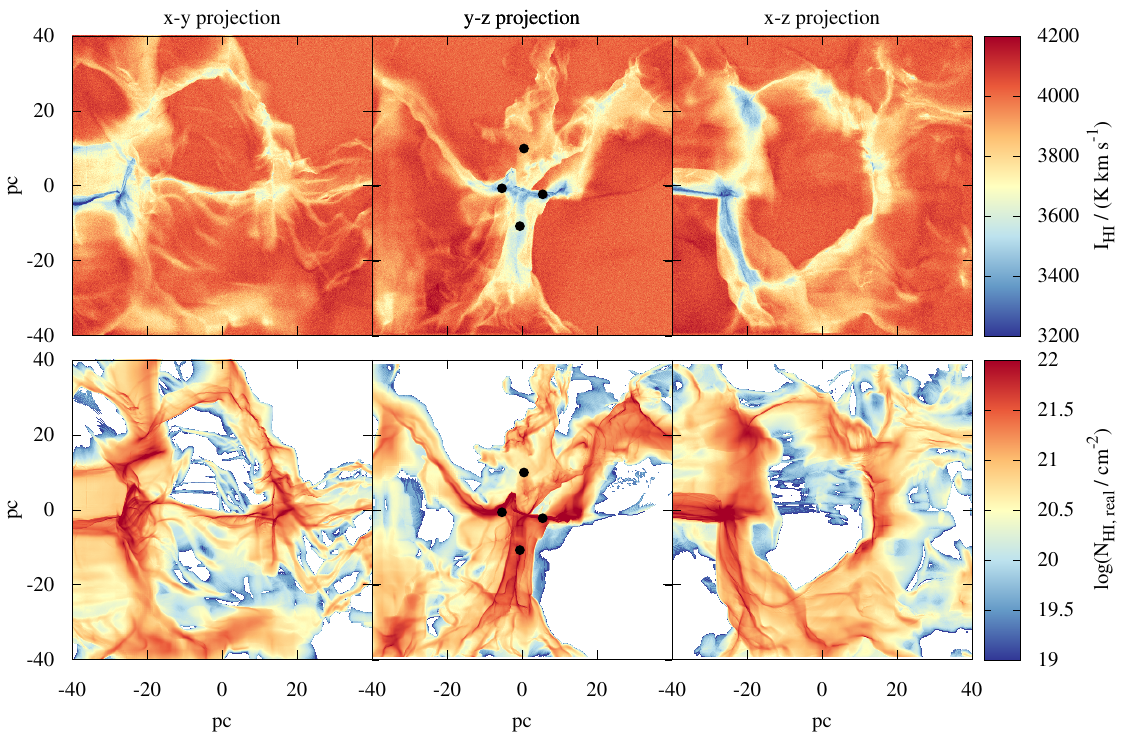} 
\caption{Top row: Integrated HI intensity for three different directions of MC1-HD at \mbox{$t_\rmn{evol}$ = 2 Myr} including observational effects for an assumed distance of 150~pc (beam size corresponds to 0.06~pc). Bottom row: Column density of HI gas with temperatures below 100~K calculated directly from the simulation data. The HISA feature traces well the high-column density regions. Note that the high integrated intensities around \mbox{4000 K km s$^{-1}$} are due to the assumed constant background brightness temperature of 100~K integrated over a velocity range of \mbox{$\pm$20 km s$^{-1}$}. The black dots in the middle column show the positions of the two LOS for which the spectra are plotted in Fig.~\ref{fig:spectrum}.}
\label{fig:HI_emission}
\end{figure*}

In order to get a first impression about the HI and H$_2$ content of the simulated MCs, in Fig.~\ref{fig:NH-diag} we show the 2D-PDFs of $N_\rmn{H_2}$, $N_\rmn{HI}$ and their ratio  vs. $N_\rmn{H, tot}$ for a selected simulation snapshot. Here, \mbox{$N_\rmn{H, tot}$ = $N_\rmn{HI}$ + 2 $N_\rmn{H_2}$ + $N_\rmn{H^+}$} denotes the total hydrogen column density. We find that H$_2$ starts to form above \mbox{$N_\rmn{H, tot}$ $\simeq$ 10$^{20}$ cm$^{-2}$}. The transition from atomic- to molecular-hydrogen dominated gas, however, occurs around \mbox{$N_\rmn{H, tot}$ $\simeq$ 10$^{21}$ cm$^{-2}$} (\mbox{$\simeq$ 8$^{}$ M$_{\sun}$ pc$^{-2}$}), as discussed in detailed in \citet{Seifried20}. This is in agreement with other numerical and semi-analytical works \citep[][]{Krumholz08,Krumholz09,Gnedin09,Sternberg14,Bialy16,Valdivia16,Bellomi20}.

However, despite H$_2$ forming rapidly above \mbox{$N_\rmn{H, tot}$ $\simeq$ 10$^{20}$ cm$^{-2}$}, also $N_\rmn{HI}$ continues to rise. Most of the HI has column densities around 10$^{21}$~cm$^{-2}$ (\mbox{$\simeq$ 8$^{}$ M$_{\sun}$ pc$^{-2}$}) similar to theoretical predictions \citep{Krumholz08,Krumholz09,Sternberg14,Bialy16}. However, even significantly higher HI column densities up to a few 10$^{22}$~cm$^{-2}$ are reached, i.e. H$_2$ and HI coexist on the projected maps. Hence, having the HI-H$_2$ transition around a certain value (e.g. 10$^{21}$~cm$^{-2}$) does \textit{not} exclude the occurrence of significantly higher HI column densities.

\subsection{Deriving $N_\rmn{HI}$ using a fixed $T_\rmn{HISA}$}
\label{sec:T_HISA_fixed}

In the following we will assess how accurately the \textit{cold} HI content in MCs can be determined via HISA observations\footnote{In \citet{Seifried20} we have already discussed a new approach to determine the H$_2$ content by means of combined CO(1-0) and dust emission observations, which allows for an accurate determination of H$_2$ within a factor  of 1.8.}. We define cold HI as all HI with temperatures below 100~K. This definition is motivated by the chosen background brightness temperature (Section~\ref{sec:HISA_calc}), which makes our synthetic HISA observations sensitive to HI with temperatures below 100~K. Depending on the simulation, we find that around and in our MCs 22 -- 43\% of the HI is warmer than 100~K, i.e. in all cases the HISA observations are sensitive to more than 50\% of the entire HI mass. In the top row of Fig.~\ref{fig:HI_emission}, we show the velocity-integrated intensity, $I_\rmn{HI}$, of our synthetic HI observations of MC1-HD from three different directions at \mbox{$t_\rmn{evol}$ = 2 Myr} including observational effects at an assumed distance of 150~pc (see Section~\ref{sec:obs_effects}). For \mbox{$t_\rmn{evol}$ = 3 Myr} the results are qualitatively similar. In the bottom row we show the cold HI column density of MC1-HD inferred directly from the simulation, $N_\rmn{HI, real}$. We emphasise that throughout this paper $N_\rmn{HI, real}$ refers to the cold HI gas (\mbox{$T$ $\leq$ 100 K}) as described above.

\begin{figure}
\includegraphics[width=0.9\linewidth]{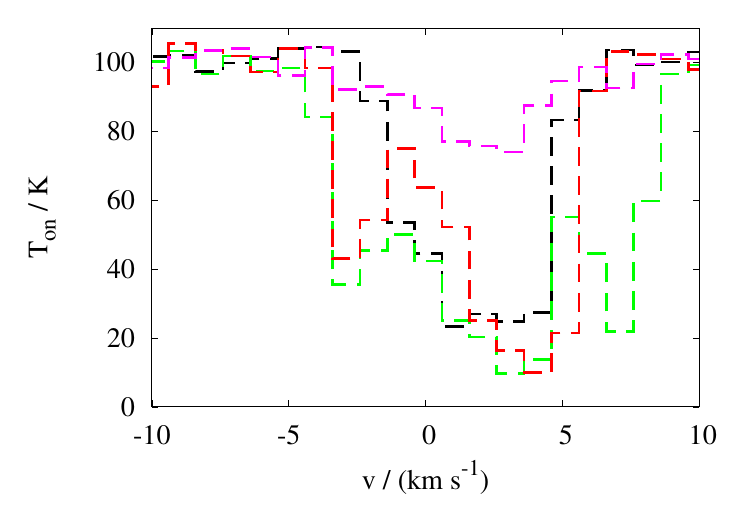} 
\caption{Synthetic HI spectra including observational effects for the positions indicated in the map in the middle column of Fig.~\ref{fig:HI_emission} (black dots) showing the variety of spectral shapes obtained in our synthetic observations. The black line shows a spectrum potentially suffering from opacity broadening at the line center. In the spectra shown by the red and green line, multiple absorption features 
seem to be present.}
\label{fig:spectrum}
\end{figure}
Both the synthetic HI emission and the $N_\rmn{HI, real}$-maps show complex filamentary structures with extended envelopes. There is a clear anti-correlation between $I_\rmn{HI}$, and $N_\rmn{HI, real}$ visible. In order to demonstrate that the drop in $I_\rmn{HI}$ is due to the self-absorption of radiation, we show in Fig.~\ref{fig:spectrum} two example spectra from pixels in the high-$N_\rmn{HI, real}$/low-$I_\rmn{HI}$ areas in the middle column of Fig.~\ref{fig:HI_emission} (black dots). For both pixels there are clear HISA features recognisable caused by the cold HI. For the other MCs and an assumed distance of 3~kpc, the obtained results are qualitatively and quantitatively very similar. We note that the rather high integrated intensities around \mbox{4000 K km s$^{-1}$} are due to the assumed constant background brightness temperature of 100~K integrated over a velocity range of \mbox{$\pm$20 km s$^{-1}$}, which is, however, automatically taken into account via Eq.~\ref{eq:Toffon}.

\begin{figure*}
\centering
\includegraphics[width=0.9\linewidth]{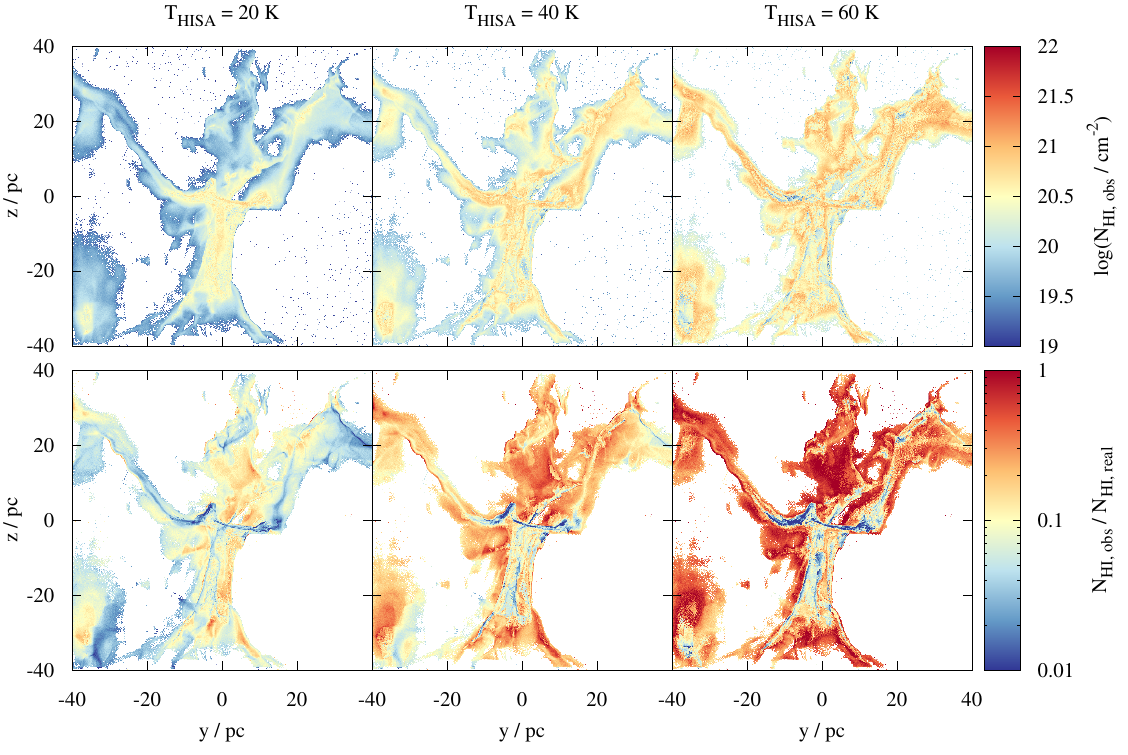} 
\caption{Top row: Observed HI column density assuming a fixed $T_\rmn{HISA}$ of 20, 40 and 60~K (from left to right) for MC1-HD at \mbox{$t_\rmn{evol}$ = 2 Myr} at an assumed distance of 150~pc (beam size corresponds to 0.06~pc). Bottom row: Ratio of the observed HI column density shown in the top row to the actual HI column density (for \mbox{$T$ $<$ 100 K}). Overall, assuming a fixed $T_\rmn{HISA}$ underestimates $N_\rmn{HI, real}$ by almost about one order of magnitude in the central, high column density regions. Increasing $T_\rmn{HISA}$ improves the match only in the outer regions.}
\label{fig:HI_map}
\end{figure*}
Next, we investigate the HI column densities obtained from the HISA observations, denoted as $N_\rmn{HI, obs}$ (Eq.~\ref{eq:NH}), assuming a fixed HISA temperature, $T_\rmn{HISA}$. In the top row of Fig.~\ref{fig:HI_map} we show $N_\rmn{HI, obs}$ for \mbox{$T_\rmn{HISA}$ = 20,} 40 and 60K for MC1-HD at \mbox{$t_\rmn{evol}$ = 2 Myr} for one LOS at an assumed distance of 150~pc. In the bottom row we show the ratio of $N_\rmn{HI, obs}$ and $N_\rmn{HI, real}$. Most prominently, we find that for all three values of $T_\rmn{HISA}$, the observed column densities $N_\rmn{HI, obs}$ in the central and most dense regions are a factor of $\gtrsim$10 too low. This trend becomes more pronounced with increasing $T_\rmn{HISA}$. In the outer regions $N_\rmn{HI, obs}$ could not be calculated for all pixels as here the absorption features are partly too weak and therefore the spectral analysis tool BTS does not identify any HISA feature. For the remaining pixels in the outer parts, $N_\rmn{HI, real}$ is also typically underestimated ranging from a few 10\% up to a factor of a few. However, increasing $T_\rmn{HISA}$ pushes $N_\rmn{HI, obs}$ in the outer regions closer to $N_\rmn{HI, real}$. Overall, however, the match between the actual and observed HI column density is rather poor with a clear tendency to underestimate $N_\rmn{HI, real}$ by factors up to $\sim$10 (see also Section~\ref{sec:totalmass} for the effect on total estimated HI mass). This does not change when considering different LOS, times, MCs or assuming a distance of 3~kpc. Reasons for this underestimation will be discussed in detail in Sections~\ref{sec:HI_temp} and~\ref{sec:optdepth}.

\begin{figure}
\centering
\includegraphics[width=0.9\linewidth]{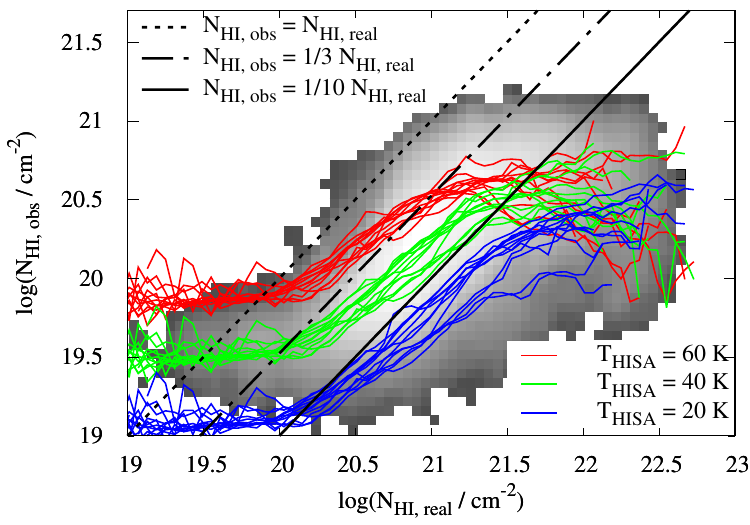} 
\caption{Mean value of $N_\rmn{HI, obs}$ against $N_\rmn{HI, real}$ for the three different directions of all four MCs placed at a distance of 150~pc at \mbox{$t_\rmn{evol}$ =  2 Myr} and using three different $T_\rmn{HISA}$ (colored lines). In the background the full distribution for one snapshot (MC1-HD, \mbox{$T_\rmn{HISA}$ = 40 K}) is shown in grey scale. The black lines show lines of constant ratio $N_\rmn{HI, obs}$/$N_\rmn{HI, real}$ to guide the readers eye. In general, the actual column density is underestimated significantly, and the maximum $N_\rmn{HI, obs}$ levels off around \mbox{10$^{21}$ cm$^{-2}$}.}
\label{fig:N_real_vs_obs}
\end{figure}
In Fig.~\ref{fig:N_real_vs_obs} we show the mean value of $N_\rmn{HI, obs}$ as a function of $N_\rmn{HI, real}$ for the four different MCs placed at 150~pc at \mbox{$t_\rmn{evol}$ = 2 Myr} focussing on the case of \mbox{$T_\rmn{HISA}$ = 20,} 40 and 60~K (coloured lines). Overall, the underestimation shown in Fig.~\ref{fig:HI_map} is observed in all cases, the degree of underestimation becomes more pronounced with increasing $N_\rmn{HI, real}$ and decreasing $T_\rmn{HISA}$. Lowering or increasing $T_\rmn{HISA}$ beyond the values shown only amplifies the trends, which is why we do not explicitly discuss them here. In the following we focus on the range of \mbox{$N_\rmn{HI, real} \geq 10^{19.5}$ cm$^{-2}$}.

First, we note that \mbox{$T_\rmn{HISA}$ = 20 K} (blue lines) is apparently a rather bad choice resulting in an underestimation by about one order of magnitude (and more). Considering \mbox{$T_\rmn{HISA}$ = 60 K} (red lines), we find that in the range \mbox{10$^{19.5}$ cm$^{-2}$ $\lesssim$ $N_\rmn{HI, real}$ $\lesssim$ 10$^{21}$ cm$^{-2}$}, the actual values are underestimated by a factor of $\sim$2 -- 5. In the same range, for \mbox{$T_\rmn{HISA}$ = 40 K} (green lines), the actual values are underestimated even more severely by a factor of $\sim$3 -- 10. Moreover, for $N_\rmn{HI, real} > 10^{21}$~cm$^{-2}$, $N_\rmn{HI, obs}$ seems to level off around $\sim$10$^{20.5 - 21}$~cm$^{-2}$, for both \mbox{$T_\rmn{HISA}$ = 40} and 60~K. This results in an increasing underestimation of the actual column density when going to denser and denser regions. We note that this artificial levelling-off around \mbox{$N_\rmn{HI, obs} \simeq 10^{21}$ cm$^{-2}$} matches well the maximum values reported in recent HISA observations \citep[][but see also Section~\ref{sec:obs} for a further discussion]{Kavars03,Kavars05,Li03,Goldsmith05,Klaassen05,Krco08,Barriault10,Krco10,Syed20,Wang20}. Finally, even for a single MC there is a large scatter of the measured $N_\rmn{HI, obs}$ for a given $N_\rmn{HI, real}$ (shown by the full distribution  in grey scale in the background for MC1-HD for \mbox{$T_\rmn{HISA}$ = 40 K}). This further lowers the accuracy with which HISA observations seem to be able to constrain the actual HI column density. An additional uncertainty in observations arises from the unknown value of $T_\rmn{off}$ (and thus $T_\rmn{off-on}$), which in our case is chosen to be constant (\mbox{$T_\rmn{off}$ = 100 K}). Eq.~\ref{eq:tau_HISA} for the optical depth implies that this uncertainty increases even further the scatter found for individual MCs at a given $N_\rmn{HI, real}$.

\subsection{The HI temperature}
\label{sec:HI_temp}

\begin{figure}
\centering
\includegraphics[width=0.9\linewidth]{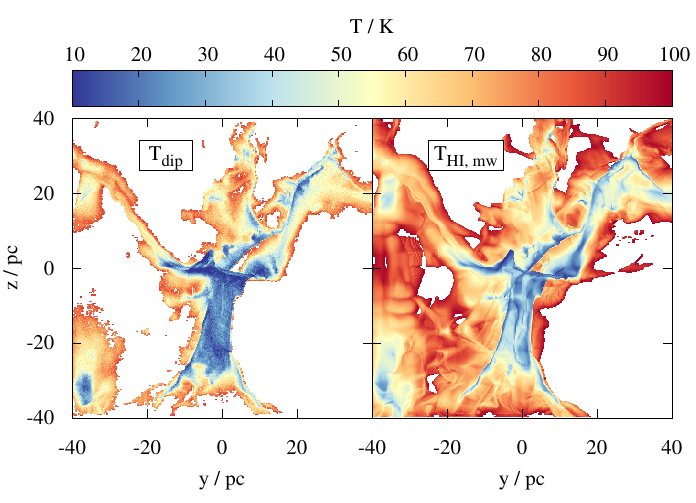} \\
\caption{Map of $T_\rmn{dip}$ (left, Eq.~\ref{eq:Tdip}), which can be used such that Eq.~\ref{eq:Tonoff} yields a result for every velocity channel, and the mass-weighted average of the temperature of the HI gas (right) for the same snapshot as shown in Fig.~\ref{fig:HI_map}. Both quantities show a reasonable agreement within about 20~K. However, $T_\rmn{dip}$  is quite low in the central areas. This explains the poor match of the observed with the actual column density (Fig.~\ref{fig:HI_map}), as here often $T_\rmn{HISA}$ $>$ $T_\rmn{dip}$ (depending on the actual choice of $T_\rmn{HISA}$ ).}
\label{fig:Tdip_map}
\end{figure}

\begin{figure}
\centering
\includegraphics[width=0.9\linewidth]{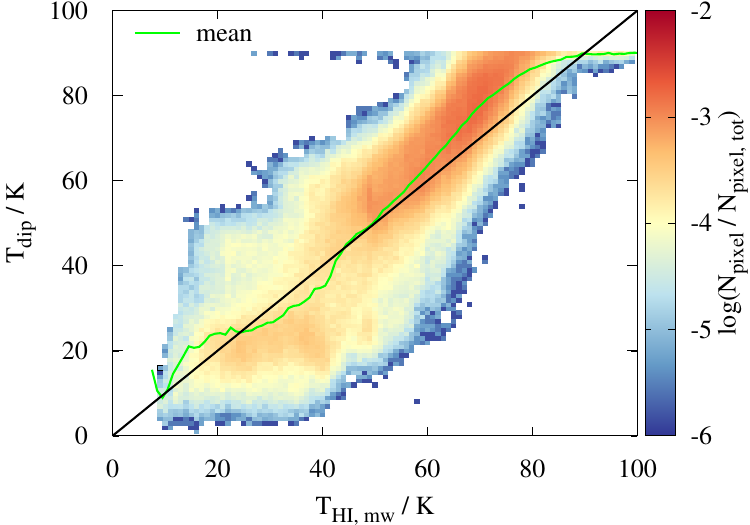} \\
\caption{2D-PDF of $T_\rmn{dip}$ vs. $T_\rmn{HI, mw}$ and its mean value (green line) for the same snapshot as shown in Fig.~\ref{fig:Tdip_map}. The black line corresponds to \mbox{$T_\rmn{dip}$ = $T_\rmn{HI, mw}$}. Overall there is a rough correspondence between both quantities with a typical scatter of $\sim$10-20~K.}
\label{fig:Tdip_THI}
\end{figure}

\begin{table*}
\centering
\caption{List of the most relevant temperature definitions used in this paper including a short explanation and the nature of each temperature.}
\begin{tabular}{p{0.05\linewidth}p{0.4\linewidth}p{0.15\linewidth}}
\hline
$T_\rmn{s}$ & \textit{actual} spin temperature of the HI gas used for the radiative transfer (Section~\ref{sec:radtrans}), close to the actual gas temperature of the simulation & spin temperature, varies along the LOS \\
\hline
$T_\rmn{HISA}$ & \textit{assumed} spin temperature of the HISA features needed for the calculation of the HI optical depth (Eq.~\ref{eq:tau_HISA}) and subsequently the column density (Eq.~\ref{eq:NH}) & spin temperature \\
\hline
$T_\rmn{HI, mw}$ & mass-weighted, LOS-averaged HI temperature \textit{calculated} from the simulation data for a given pixel, see right panel of Fig.~\ref{fig:Tdip_map} & kinetic gas temperature \\
\hline
$T_\rmn{HI, min}$ & minimum HI temperature \textit{calculated} from the simulation data along the LOS for a given pixel & kinetic gas temperature \\
\hline
$T_\rmn{on}(v)$ & spectrum of the \textit{measured} HI brightness temperature for a given pixel, see Fig.~\ref{fig:spectrum} & brightness temperature \\
\hline
$T_\rmn{dip}$ & lowest temperature of the $T_\rmn{on}(v)$-spectrum for a given pixel, see Eq.~\ref{eq:Tdip} and left panel of Fig.~\ref{fig:Tdip_map} & brightness temperature\\
\hline
\end{tabular}
\label{tab:temp}
\end{table*}

Investigating Eq.~\ref{eq:Tonoff} shows that there is a certain upper threshold for $T_\rmn{HISA}$, above which the equation is not solvable for $\tau_\rmn{HISA}$ for at least some of the velocity channels, which then would have to be omitted for the calculation of $N_\rmn{HI}$.  We denote this upper threshold as $T_\rmn{dip}$, which is set by the minimum of $T_\rmn{HISA,max}(v)$ (Eq.~\ref{eq:Tmax}) over all velocity channels for a given pixel, i.e.
\begin{equation}
 T_\rmn{dip} = \text{min}(T_\rmn{HISA, max}(v))  = \text{min}(T_\rmn{on}(v)) \, .
 \label{eq:Tdip}
\end{equation}
The denomination as $T_\rmn{dip}$ is motivated by the fact that it corresponds to the temperature at the dip of the observed absorption spectrum $T_\rmn{on}(v)$. For the sake of clarity, in Table~\ref{tab:temp} we give a short summary of the most relevant temperature definitions used in this paper.

\begin{figure}
\centering
\includegraphics[width=\linewidth]{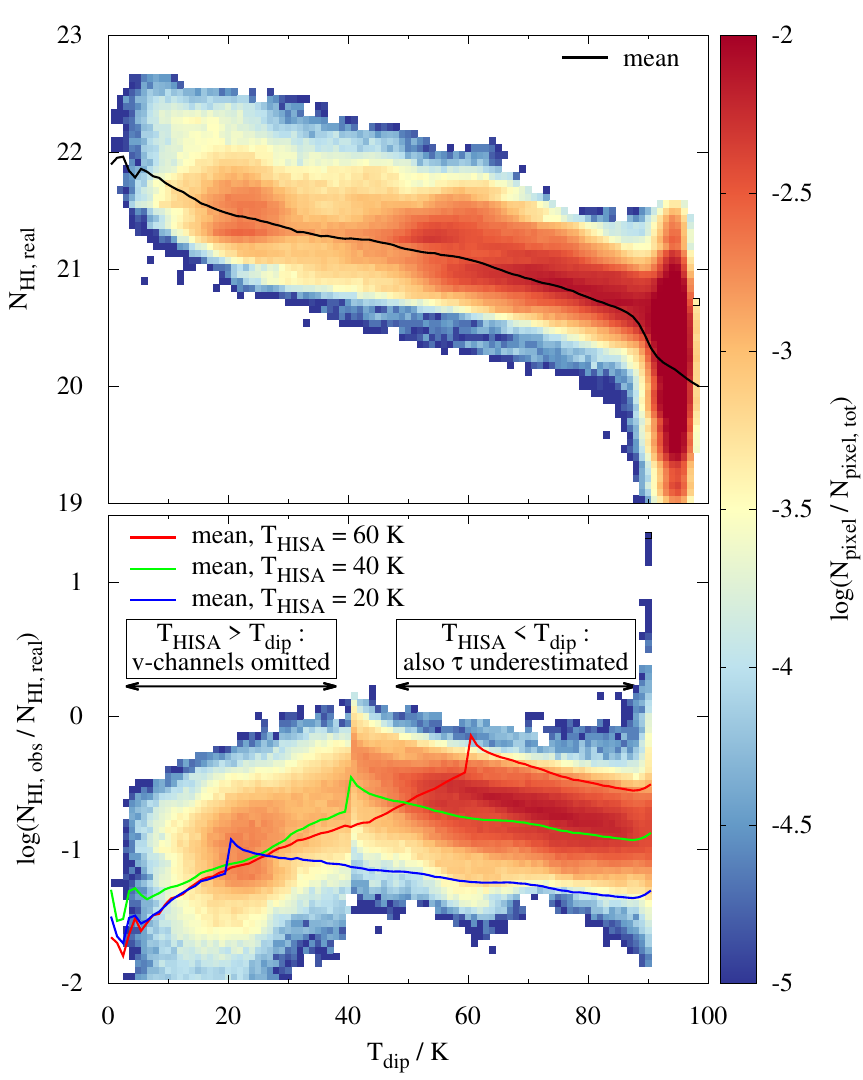} 
\caption{Top: Phase diagram of $N_\rmn{HI, real}$ vs. $T_\rmn{dip}$ for the same snapshot as in Fig.~\ref{fig:HI_map}. The highest HI column densities are associated with low $T_\rmn{dip}$. Bottom: Phase diagram of the ratio of $N_\rmn{HI, obs}$ and $N_\rmn{HI, real}$ vs. $T_\rmn{dip}$ for the same snapshot as in the top panel using \mbox{$T_\rmn{HISA}$ = 40 K}. The coloured lines show the mean value of the distribution for three different $T_\rmn{HISA}$. Overall the actual HI column density is underestimated by up to a factor of $\sim$10. Changing $T_\rmn{HISA}$ increases the accuracy only locally, i.e. where that assumed $T_\rmn{HISA}$ roughly corresponds to $T_\rmn{dip}$, which represents the real temperature of the HISA feature. We indicated the reasons of the underestimation in the two temperature ranges above and below \mbox{$T_\rmn{dip}$ = $T_\rmn{HISA}$}.}
\label{fig:Tdip_Nratio}
\end{figure}

In the left panel of Fig.~\ref{fig:Tdip_map} we show the map of $T_\rmn{dip}$ for one snapshot. In addition, the right panel shows the actual mass-weighted, LOS-averaged HI temperature, $T_\rmn{HI,mw}$\footnote{As the synthetic HI observations are only sensitive to HI with temperatures below 100~K, also for this average (as for $N_\rmn{HI, real}$) only HI gas with $T$~$\leq$~100~K is considered.}. Both $T_\rmn{dip}$ and $T_\rmn{HI,mw}$ show a strong drop towards the central, high-column density areas (see also top panel of Fig.~\ref{fig:Tdip_Nratio}). Interestingly, we find that both temperature measures show an agreement within about 20~K (Fig.~\ref{fig:Tdip_THI}). Based on this, in Section~\ref{sec:T_HISA_free} we investigate whether the usage of $T_\rmn{dip}$ as an approximation for the temperature of the cold HI along the LOS (and thus for $T_\rmn{HISA}$) is suitable.

The strong variations of $T_\rmn{dip}$ and $T_\rmn{HI, mw}$ down to values as low as $\sim$10~K cause the significant underestimation of $N_\rmn{HI, real}$ seen in the Figs.~\ref{fig:HI_map} and~\ref{fig:N_real_vs_obs} by means of two effects explained in the following and sketched in Fig.~\ref{fig:Tdip_Nratio}:
\begin{enumerate}
\item \mbox{$T_\rmn{HISA}$ chosen too high ($>$ $T_\rmn{dip}$)}: These regions typically correspond to high $N_\rmn{HI, real}$ ($\gtrsim10^{21}$~cm$^{-2}$) (top panel of Fig.~\ref{fig:Tdip_Nratio}). Here, Eq.~\ref{eq:Tonoff} does not yield any results for at least some of the velocity channels. This happens for $\lesssim$10\%, 5 - 20\%, and 30 - 40\% of the pixels for $T_\rmn{HISA}$~= 20, 40, and 60~K, respectively. Hence, for these pixels \textit{some} of the channels have to be neglected, which reduces $N_\rmn{HI, obs}$ significantly (Eq.~\ref{eq:NH}). As with decreasing $T_\rmn{dip}$, i.e. increasing $N_\rmn{HI, real}$, more and more velocity channels have to be omitted (at a fixed $T_\rmn{HISA}$), this leads to the observed artificial levelling-off of $N_\rmn{HI, obs}$ at $\sim$10$^{21}$~cm$^{-2}$. In consequence, for central, high column density regions of MCs, $N_\rmn{HI, real}$ is underestimated by a factor of about 10 and more (Fig.~\ref{fig:N_real_vs_obs}). 
\item \mbox{$T_\rmn{HISA}$chosen too low ($<$ $T_\rmn{s}$)}:
This leads to an underestimation of the optical depth $\tau_\rmn{HISA}$ (Eq.~\ref{eq:tau_HISA}) as well as $N_\rmn{HI, obs}$ (Eq.~\ref{eq:NH}, both via $\tau_\rmn{HISA}$ and  the assumption \mbox{$T_\rmn{s} = T_\rmn{HISA}$}). This effect is dominant mainly in the low to intermediate column density regions in the outer parts of the MCs (\mbox{$N_\rmn{HI, real}$ $\lesssim$ 10$^{21}$ cm$^{-2}$}). Here, $N_\rmn{HI, obs}$ underestimates $N_\rmn{HI, real}$ on average by a factor of 3 -- 10. Due to the linear dependence of $N_\rmn{HI, obs}$ on $T_\rmn{HISA}$, the actual value of $N_\rmn{HI, obs}/N_\rmn{HI, real}$ increases with increasing $T_\rmn{HISA}$ at high $T_\rmn{dip}$ (coloured lines in the bottom panel of Fig.~\ref{fig:Tdip_Nratio}).
\end{enumerate}
Overall, our results demonstrate that finding an accurate value for $T_\rmn{HISA}$ is crucial but at the same time \textit{not possible} when using a single value for the entire map. In addition, both choosing a too high or too low value for $T_\rmn{HISA}$ leads to an underestimation of the HI column density.

\subsubsection{HI temperature variations}

Interestingly,  even for regions where \mbox{$T_\rmn{HISA}$ $\simeq$ $T_\rmn{dip}$} (indicated by  the peak of the coloured lines in the bottom panel of Fig.~\ref{fig:Tdip_Nratio}), a ratio of $N_\rmn{HI, obs}/N_\rmn{HI, real}$ close to 1 is barely reached. An additional source of uncertainty causing this are the significant variations of $T_\rmn{HI,mw}$ across the map (right panel of Fig.~\ref{fig:Tdip_map}). Similar variations will also occur for each individual pixel \textit{along} the LOS. Hence, even for an individual pixel the assumption of a constant $T_\rmn{HISA}$ presents an oversimplification, which in turn results in the observed underprediction of the HI column density particularly for the dense regions. 

\begin{figure}
\centering
\includegraphics[width=0.9\linewidth]{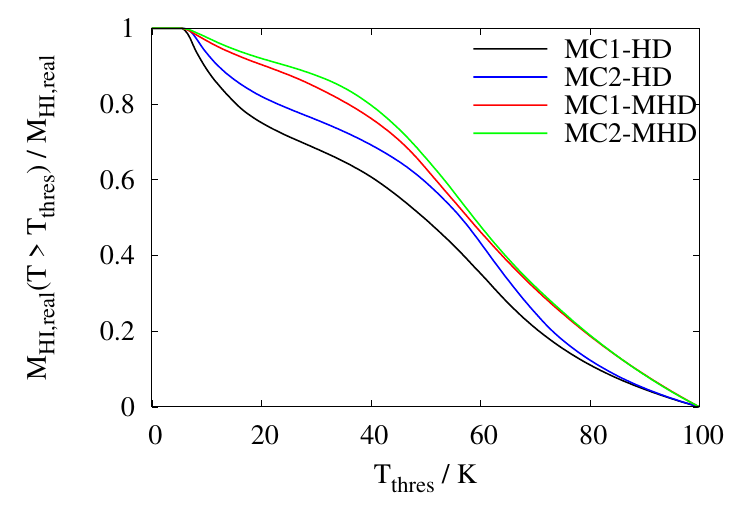} 
\caption{Cumulative temperature PDF showing the amount of cold HI above a certain threshold temperature for all four MCs at \mbox{$t_\rmn{evol}$ = 2 Myr}. The steady rise with decreasing temperature indicates that no single temperature choice for $T_\rmn{HISA}$ is suitable to accurately determine the amount of cold HI in the clouds. Note that gas above 100~K is not considered here.}
\label{fig:HI_temp}
\end{figure}

We emphasise that $T_\rmn{HI,mw}$ can be a few 10~K higher than the temperature of the coldest HI gas along each LOS, denoted as $T_\rmn{HI, min}$  (not shown), due to HI gas warmer than $T_\rmn{HI, min}$ along the LOS. This also explains why there are regions where $T_\rmn{dip} < T_\rmn{HI,mw}$ (Fig.~\ref{fig:Tdip_THI}), whereas $T_\rmn{HI, min}$ is, as expected, always smaller than $T_\rmn{dip}$. This also indicates that $T_\rmn{dip}$ only gives an upper limit to the actual spin temperature $T_\rmn{s}$ of the absorbing HI layer. The non-isothermality of the cold HI gas is further emphasised by its temperature distribution in Fig.~\ref{fig:HI_temp} showing the cumulative PDF of HI gas above a certain threshold temperature for all four MCs at \mbox{$t_\rmn{evol}$ = 2 Myr}. The amount of HI gas is rising steadily with decreasing temperature, independent of the considered MC. This shows that no \textit{single} temperature can be used to describe the HI content of MCs. As an example, for  \mbox{$T_\rmn{HISA}$ = 40 K}, the HISA observations would (at least) miss out 20 -- 40\% of the cold HI mass.

The above results explain why also in the absence of observational effects like noise and limited spectral or spatial resolution the poor match between the observed and actual HI column density remains (see Fig.~\ref{fig:HI_map_no_noise} in Appendix~\ref{sec:appendixa}). Also for different assumed distances of 150~pc and 3~kpc we find little differences. This further supports our claim that the underestimation of $N_\rmn{HI, real}$ can be attributed -- at least in parts (see Section~\ref{sec:optdepth}) -- to the non-uniform HI temperatures present in the clouds.

Finally, we note that the rather low HI temperatures found in our simulations ($\lesssim$~40~K, Fig.~\ref{fig:Tdip_map}) are in good agreement with a number of observations of Galactic MCs, which find typical HI temperatures between 10~K and 40~K \citep{Gibson00,Kavars03,Kavars05,Klaassen05,Fukui14,Fukui15,Stanimirovic14,Denes18,Nguyen19}. They are, however, lower than typical temperatures found by \citet[][]{Wang20} in the giant molecular filament GMF38a. A possible reason for this might be stellar feedback heating the gas in GMF38a. In consequence, the HI column densities determined in \citet{Wang20} might be more accurate than in our case.

\subsection{The HI optical depth}
\label{sec:optdepth}

\begin{figure}
\centering
\includegraphics[width=0.9\linewidth]{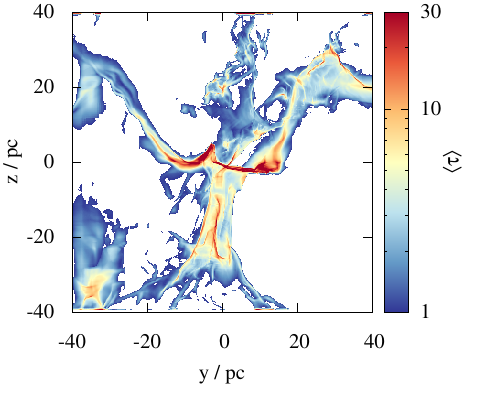} 
\caption{Map of the HI optical depth proxy $\langle \tau \rangle$ (see Appendix~\ref{sec:appendixb}) for MC1-HD at 2~Myr. The dense cloud region (compare with bottom middle panel of Fig.~\ref{fig:HI_emission}) has an average optical depth $\gtrsim$ 1, thus optical depth corrections cannot be neglected when calculating HI column densities.} 
\label{fig:tau}
\end{figure}
In order to investigate the typical optical depths in our clouds, in Fig.~\ref{fig:tau} we plot a proxy for the HI optical depth, $\langle \tau \rangle$, for MC1-HD at 2~Myr. The definition of $\langle \tau \rangle$ is given in Appendix~\ref{sec:appendixb}. It represents a channel-averaged approximation to the real optical depth which is accurate within a few 10\% above $\langle \tau \rangle$ = 1, i.e. in the optically thick regions we are interested in here. For optically thin regions, the approximation is not applicable, which is why we do not show these regions here. The values of $\langle \tau \rangle$ span a wide range, from the moderately optically thick regime up to highly optically thick regions with $\langle \tau \rangle$ $\sim$ 10. In particular, the entire area of central cloud (compare with bottom middle panel of Fig.~\ref{fig:HI_emission}) has an optical depth \mbox{$\gtrsim$ 1}, which is in excellent agreement with recent observations \citep[e.g.][]{Fukui14,Fukui15,Bihr15,Denes18,Murray18,Nguyen19,Syed20,Wang20}. Our results thus demonstrates that optical depth effects cannot be neglected in HI observations of MCs, also when calculating $N_\rmn{HI}$ from HI emission observations.

We note that at first view the maximum values of \mbox{$\langle \tau \rangle > 10$} appear high in comparison with those found in the aforementioned observational works. However, as these measurements are limited by observational noise, $\Delta T$, the observationally reported values have to be taken as lower limits \citep[see e.g. fig.~10 of][]{Bihr15}.

In regions of high optical depth, the observed brightness temperature $T_\rmn{on}$ will be close to the spin temperature $T_\rmn{s}$ of the absorbing, cold HI layer, i.e.  \mbox{$T_\rmn{off-on}$ $\simeq$ 100 K - $T_\rmn{s}$}. Hence, choosing $T_\rmn{HISA} \lessgtr T_\rmn{s}$ will result in an underestimation of $N_\rmn{HI, real}$ for optically thick regions as well, as discussed in Section~\ref{sec:HI_temp}. Moreover, for $T_\rmn{HISA} \simeq T_\rmn{s}$ an additional source of error in such optically thick regions is caused by the observational noise $\Delta T$, as now \mbox{$T_\rmn{off-on}$ $\simeq$} \mbox{100 K - $T_\rmn{HISA} - \Delta T$} (from Eq.~\ref{eq:Toffon}). Inserting this into Eq.~\ref{eq:tau_HISA} yields
\begin{equation}
\tau_\rmn{HISA, noise} = -\textrm{ln} \left( 1 - \frac{100 \, \rmn{K} - T_\rmn{HISA} - \Delta T}{100 \, \rmn{K}  - T_\rmn{HISA}} \right) \, = -\textrm{ln}\left(\frac{\Delta T}{100 \, \rmn{K}  - T_\rmn{HISA}} \right) \, .
\label{eq:deltatau}
\end{equation}
Analysing Eq.~\ref{eq:deltatau} shows that the observational uncertainty $\Delta T$ in highly optically thick regions (if $T_\rmn{HISA} \simeq T_\rmn{s}$) results in an underestimation of $N_\rmn{HI, obs}$ \textit{regardless} of its sign:
\begin{enumerate}
\item $\Delta T < 0$: If noise artificially lowers $T_\rmn{on}$, this \textit{increases} $T_\rmn{off-on}$ beyond a value of \mbox{100 K - $T_\rmn{HISA}$} in a highly optically thick region. Hence, Eq.~\ref{eq:deltatau} would contain a negative expression in the logarithm and the contribution from the corresponding velocity channel has to be omitted.
\item $\Delta T > 0$: If noise, but also the potential emission of warm and diffuse HI in the foreground, increases $T_\rmn{on}$ (and thus decreases $T_\rmn{on-off}$), this results in an underestimation of the true value of $\tau_\rmn{HISA}$ (which can be larger than $\tau_\rmn{HISA, noise}$) and thus also $N_\rmn{HI, obs}$ (Eq.~\ref{eq:NH}). The  effect of foreground emission is thus also related to the problem of identifying $T_\rmn{HISA}$ correctly.
\end{enumerate}
Hence, even if one were to choose the correct value of $T_\rmn{HISA}$, $N_\rmn{HI, obs}$ is in general underestimated in optically thick regions (see regions of high $N_\rmn{HI, real}$ in Fig.~\ref{fig:N_real_vs_obs}). The observational noise contributes to the fact that even at the peaks of the mean-value lines in the bottom panel of Fig.~\ref{fig:Tdip_Nratio}, where $T_\rmn{HISA}$~$\simeq$~$T_\rmn{dip}$, the real HI column density is on average underestimated. We emphasise that this underestimation due to $\Delta T$ adds on top of the problem to determine a reasonable value of $T_\rmn{HISA}$. This effect also contributes to the more pronounced underestimation at low values of $T_\rmn{dip}$ (higher values of $N_\rmn{HI, real}$, bottom panel of Fig.~\ref{fig:Tdip_Nratio}): the lower $T_\rmn{dip}$, the higher is the HI column density and thus the optical depth, which amplifies the issue arising from this effect.

\subsubsection{Opacity correction in MCs}
\label{sec:opacity_correction}

Motivated by the large extent of optically thick regions in MCs (Fig.~\ref{fig:tau}), we suggest a method to improve the accuracy of $N_\rmn{HI, obs}$: A significant underestimation occurs in the high-N$_\rmn{HI}$/high-optical depth regions of the MCs, where Eq.~\ref{eq:tau_HISA} yields no result for $\tau_\rmn{HISA}$ any more and velocity channels have to be omitted (see Fig.~\ref{fig:Tdip_Nratio}). Hence, for these velocity channels we use an optical depth set by the typical rms noise ($\Delta T$) of the observation, which is given by $\tau_\rmn{HISA, noise}$ (Eq.~\ref{eq:deltatau}), e.g. for \mbox{$T_\rmn{HISA}$ = 40 K} and the adopted noise of 3~K (Section~\ref{sec:obs_effects}), we obtain \mbox{$\tau_\rmn{HISA, noise}$ = 3.0}. We emphasise that this estimate is still a conservative estimate as the actual optical depth is likely to be higher.  A similar approach is also followed by \citet{Bihr15} for HI emission maps. We note that this approach has to be considered under the premise that, as shown before, a constant value of $T_\rmn{HISA}$ over the entire map is an oversimplification, which in addition does also not account for the temperature variations along the LOS (see Section~\ref{sec:HI_temp}).

\begin{figure}
\centering
\includegraphics[width=0.9\linewidth]{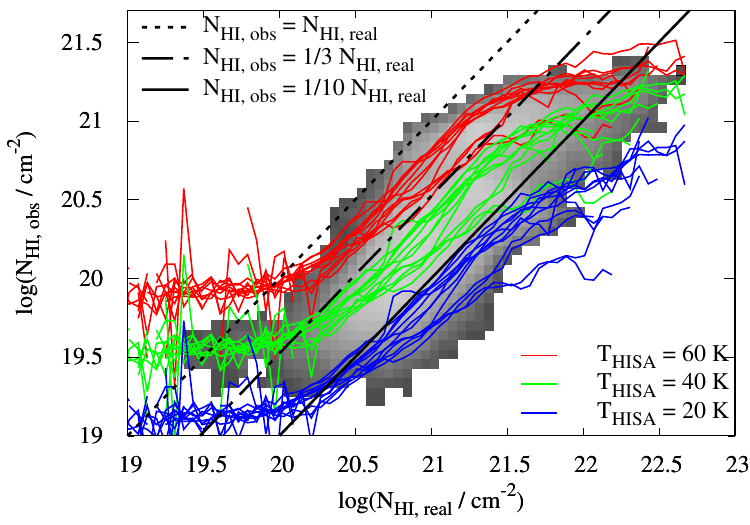} 
\caption{Same as in Fig.~\ref{fig:N_real_vs_obs} but now including the correction in optically thick regions (see text). Overall, the match is somewhat improved. However, depending on the choice of $T_\rmn{HISA}$, $N_\rmn{HI, real}$ is still underestimated by a factor of a few up to $\sim$10 in particular in the densest regions.}
\label{fig:N_real_vs_obs_added}
\end{figure}
The obtained column density maps are shown in Fig.~\ref{fig:HI_map_added} in Appendix~\ref{sec:appendixa}. As can be seen, $N_\rmn{HI, obs}$ in the denser parts of the MCs is represented better than before (compare with Fig.~\ref{fig:HI_map}). In the very densest parts, however, $N_\rmn{HI, real}$ is still significantly underestimated. This is also visible in Fig.~\ref{fig:N_real_vs_obs_added}, where we show the mean values of $N_\rmn{HI, obs}$ for all MCs and directions at 2~Myr using this correction. There is an improvement in all areas compared to the case without any correction (see Fig.~\ref{fig:N_real_vs_obs}). However, $N_\rmn{HI, real}$ can still be underestimated by a factor of a few to $\sim$10. Hence, the suggested method has only a moderate impact on increasing the accuracy, both due to the non-isothermality of the HI gas and the fact that $\tau_\rmn{HISA, noise}$ is most likely lower than the real optical depth.

\subsection{The cold HI budget of molecular clouds}
\label{sec:totalmass}

Summarizing the findings of the previous sections we find that the uncertainty in determining the HI column density is due to (i) the assumption of a fixed temperature $T_\rmn{HISA}$ for the calculation of the HISA column densities and (ii) noise in the temperature brightness measurement. As a consequence, either the optical depth and the true $T_\rmn{HISA}$ are underestimated (mainly in the outer parts of MCs) or velocity channels have to be omitted for the calculation of the column density (mainly in the densest parts of MCs). Overall, this results in a significant underestimation of the actual HI column densities by a factor of 3 - 10 (and even more in the densest regions of clouds).

\begin{figure}
\centering
\includegraphics[width=\linewidth]{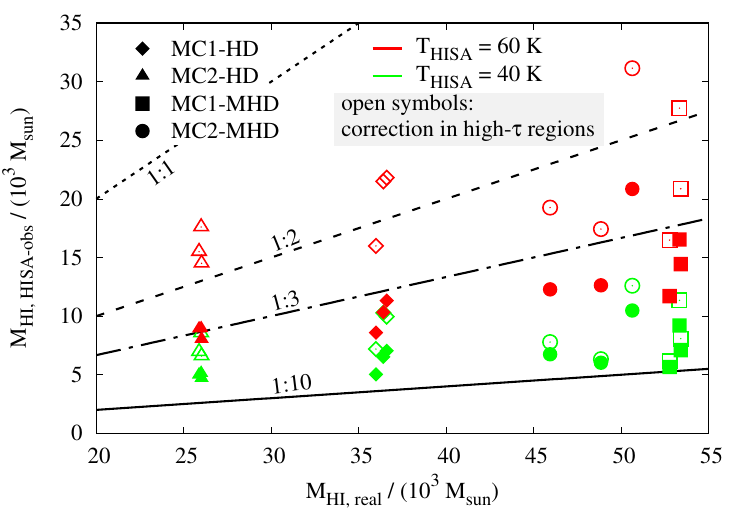} 
\caption{Accuracy of the HI mass inferred from HISA observations for the four different MCs at 2~Myr (symbols) for \mbox{$T_\rmn{HISA}$ = 40 K (green)} and 60~K (red). The black lines show the different mass ratios to guide the readers eye. Overall, we find that the HI mass is underestimated by a factor of a few up to $\sim$10. Correction for optically thick channels (open symbols) improves the accuracy only moderately. Note that depending on the chosen projection direction, the mass in the observable area is different for the same MC.}
\label{fig:masses}
\end{figure}
This is also reflected in the total mass of cold HI in MCs inferred from HISA observations, $M_\rmn{HI, HISA-obs}$, shown in Fig.~\ref{fig:masses}. Here, we add up the observed  HI mass of all pixels for which a HISA feature is identified (i.e. pixels outside the coloured regions in Fig.~\ref{fig:HI_map} are ignored) using \mbox{$T_\rmn{HISA}$ = 40 K} and 60~K. For the actual mass, $M_\rmn{HI, real}$ , we only take into account HI gas with \mbox{$T$ $<$ 100 K}, to which our HISA observations are sensitive to. As for the column densities, also the total, cold HI mass is typically underestimated by a factor of a 3 -- 10, when no correction in the optically thick regions is applied (filled symbols). The correction (Section~\ref{sec:opacity_correction}), however, improves the accuracy only moderately by a factor of \mbox{$\sim$1.5 - 2} (open symbols). We emphasise that increasing $T_\rmn{HISA}$ to obtain apparently more accurate mass estimates should be considered with caution. This merely leads to an overestimation of $N_\rmn{HI}$ at low column densities compensating the underestimation at high column densities (see Fig.~\ref{fig:N_real_vs_obs} and also Section~\ref{sec:T_HISA_free}).

We emphasise that our results do not change significantly among the different MCs considered, i.e. whether or not dynamically important magnetic fields are present. This indicates that HISA observations in general tend to significantly underestimate the cold HI budget in MCs. This is markedly different to a complementary study for the more diffuse ISM (on scales \mbox{$\gtrsim$ 1pc}) of \citet{Kim14} and \citet{Murray15,Murray17}, who find that in this regime HI absorption observations can trace the HI mass with an accuracy of a few 10\%. We tentatively attribute this to the fact that the authors probe several and more diffuse HI clouds along significantly longer LOS of several 100~pc length. These clouds might have lower optical depths and are thus less prone to the measurement uncertainties mentioned in Section~\ref{sec:optdepth}. Furthermore, HI masses obtained from emission observations \citep[e.g.][]{Bihr15}, which correct for the optical depth and which also implicitly take into account warm HI (\mbox{$T >$ 100~K}, in our case \mbox{22 -- 47\%}), might achieve more accurate HI masses, a topic not investigated in this study.

\subsection{Deriving $N_\rmn{HI}$ with a variable $T_\rmn{HISA}$}
\label{sec:T_HISA_free}

\subsubsection{$T_\rmn{HISA}$ as a free fit parameter}
\label{sec:T_HISA_Knapp}

\begin{figure*}
\includegraphics[width=\textwidth]{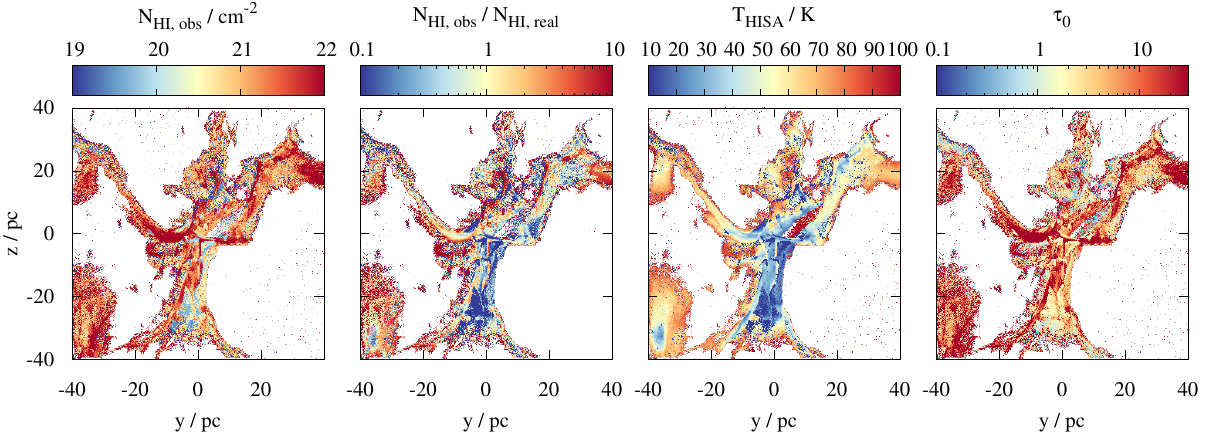} 
\caption{Maps of $N_\rmn{HI, obs}$ and its ratio to the actual HI column density (two left panels) determined by leaving $T_\rmn{HISA}$ and $\tau_0$ (two right panels) as free parameters. The maps are shown for the same snapshot as in Fig.~\ref{fig:HI_map}. Overall, allowing $T_\rmn{HISA}$ to be a free parameter does not increase the accuracy compared to assuming a fixed $T_\rmn{HISA}$ (compare Fig.~\ref{fig:HI_map}). Although the determined $T_\rmn{HISA}$ is similar to the mass-weigthed HI temperature (right panel of Fig.~\ref{fig:Tdip_map}), the optical depth shows a partly unphysical behaviour with high values in the outer, low-column density areas.}
\label{fig:HI_Knapp}
\end{figure*}

\begin{figure}
\centering
\includegraphics[width=0.9\linewidth]{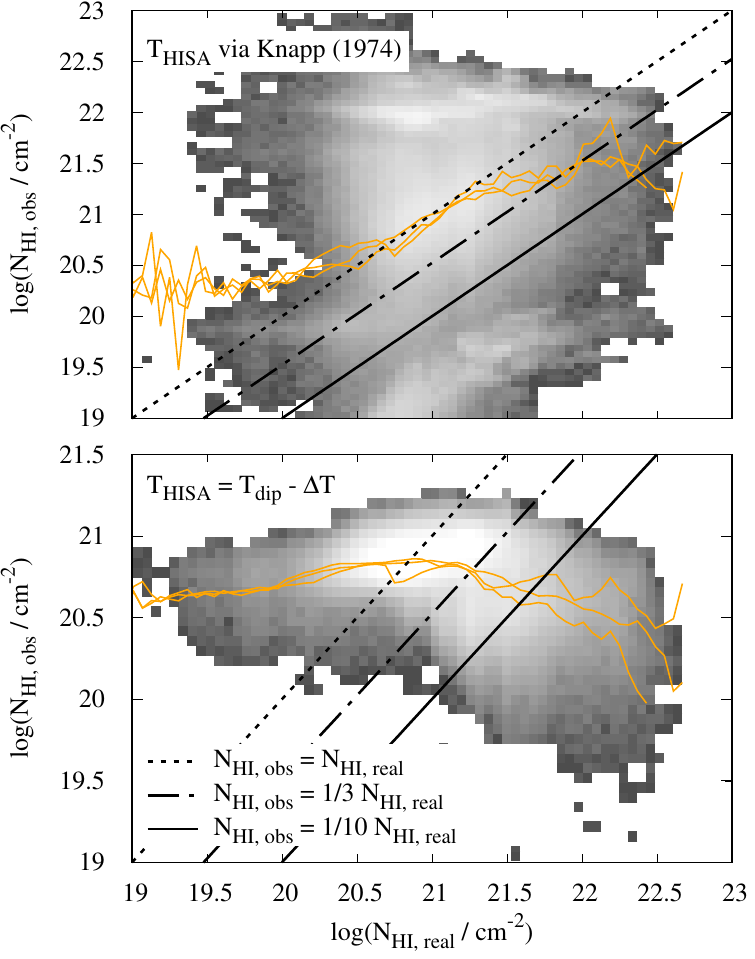}
\caption{Mean value of $N_\rmn{HI, obs}$ against $N_\rmn{HI, real}$ (orange lines) for three different directions of MC1-HD at \mbox{$t_\rmn{evol}$ =  2 Myr} using the method of \citet{Knapp74} (top) and Eq.~\ref{eq:THISATdip} with \mbox{$\Delta T$ = 3 K} (bottom). In the background, the full distribution for one LOS is shown in grey scale. The black lines show lines of constant ratio $N_\rmn{HI, obs}$/$N_\rmn{HI, real}$ to guide the readers eye. In general, the qualitative match between $N_\rmn{HI, obs}$ and $N_\rmn{HI, real}$ is rather poor with a significant scatter for the \citet{Knapp74} method. Note the different ranges on the y-axis.}
\label{fig:Nobs_TKnapp_Tdip}
\end{figure}

Following the results of Section~\ref{sec:HI_temp}, we next leave $T_\rmn{HISA}$ as a free parameter. We determine its value and the optical depth by assuming a Gaussian optical depth profile, i.e. inserting Eq.~\ref{eq:tau_Knapp} in Eq.~\ref{eq:Toffon} and fitting the observed HI spectrum. In Fig.~\ref{fig:HI_Knapp} we show the various quantities obtained by the approach for MC1-HD at 2~Myr and an assumed distance of 150~pc. For other directions, times, clouds, and the 3~kpc-distance case, we find qualitatively and quantitatively similar results. Overall, we find a quite poor match between the observed and actual $N_\rmn{HI}$ (second panel from the left in Fig.~\ref{fig:HI_Knapp} and top panel of Fig.~\ref{fig:Nobs_TKnapp_Tdip}): Although the mean values of the distribution (orange lines in the top panel of Fig.~\ref{fig:Nobs_TKnapp_Tdip}) show a reasonable match for $N_\rmn{HI, obs}$ and $N_\rmn{HI, real}$ below $\sim10^{21.5}$~cm$^{-2}$, there is a significant scatter of more than 1~dex.

We attribute this rather poor match mainly to (i) the occurrence of multiple Gaussian absorption components in the spectra (see Fig.~\ref{fig:spectrum}), which are not accounted for in our simplistic model, and  -- to a lesser extent -- to (ii) the lack of spectral resolution (\mbox{1 km s$^{-1}$}) and (iii) observational noise. In consequence, it is not possible to reliably determine the optical depth with our fitting approach, which would required accurate spectral information, also about the wings of the spectrum. This is visible in the obtained values of $\tau_0$ (right-most panel in Fig.~\ref{fig:HI_Knapp}), which show no clear correlation with the underlying column density distribution $N_\rmn{HI, real}$ as opposed to the optical depth proxy $\langle \tau \rangle$ shown in Fig.~\ref{fig:tau}.

We emphasise that when repeating the method for the noiseless, high-resolution spectra (\mbox{200 m s$^{-1}$}), we obtain a similar poor match between $N_\rmn{HI, obs}$ and $N_\rmn{HI, real}$. This further supports our assumption that the poor match is in parts due to the occurrence of multiple Gaussian absorption components not accounted for here and not due to a generic problem of this approach. We therefore suggest that multiple Gaussian components with individual temperatures have to be taken into account \citep[e.g.][]{Heiles03a,Stanimirovic14, Murray15,Denes18} to get a better match with $N_\rmn{HI, real}$. This might, to some extent also remedy the temperature problem discussed in Section~\ref{sec:HI_temp} as each component can be assigned an individual temperature. We will, however, postpone this investigation to future work.

Finally, we note that the fitted values of $T_\rmn{HISA}$ (second panel from the right in Fig.~\ref{fig:HI_Knapp}) appear roughly comparable to the mass-weighted mean temperatures (right panel of Fig.~\ref{fig:Tdip_map}). However, also here strong temperature variations along the LOS (Section~\ref{sec:HI_temp}) can affect the fit value of $T_\rmn{HISA}$. In consequence, as $\tau_0$ and $T_\rmn{HISA}$ are degenerate, overestimating (underestimating) $T_\rmn{HISA}$ requires a higher (lower) $\tau_0$ to match the observed $T_\rmn{off-on}$ at the dip of the absorption spectrum (Eq.~\ref{eq:Toffon}). Following Eq.~\ref{eq:NH}, this directly leads to a too high (low) value of $N_\rmn{HI, obs}$.

\subsubsection{$T_\rmn{HISA}$ given by $T_\rmn{dip}$}

Given the similarity of $T_\rmn{dip}$ and the mass-weighted HI temperature $T_\rmn{HI, mw}$ (see Fig.~\ref{fig:Tdip_THI}), as well as the fact that the highest accuracy for $N_\rmn{HI, obs}$ was found where $T_\rmn{HISA} \simeq T_\rmn{dip}$ (bottom panel of Fig.~\ref{fig:Tdip_Nratio}), we also try an alternative approach by setting $T_\rmn{HISA}$ close to, but slightly below $T_\rmn{dip}$. In detail, we set
\begin{equation}
 T_\rmn{HISA} = T_\rmn{dip} - \Delta T
 \label{eq:THISATdip}
\end{equation}
with \mbox{$\Delta T$ = 3 K} being the noise level of the synthetic observations and test the approach for MC1 at \mbox{$t_\rmn{evol}$ = 2 Myr}.

As for the method from \citet{Knapp74}, we find a qualitatively poor match between $N_\rmn{HI, obs}$ and $N_\rmn{HI, real}$ (bottom panel of Fig.~\ref{fig:Nobs_TKnapp_Tdip}). The method gives a rather flat distribution of $N_\rmn{HI, obs}$ with too high values at low $N_\rmn{HI, real}$ and drops towards higher $N_\rmn{HI, real}$. Overall, the reasons for this poor match are again the temperature variations along the LOS (Section~\ref{sec:HI_temp}), the partly high optical depths (for the high-$N_\rmn{HI}$ regions, Section~\ref{sec:optdepth}) as well as the degeneracy of $T_\rmn{HISA}$ and $\tau_\rmn{HISA}$ (for the low $N_\rmn{HI}$ regions, Section~\ref{sec:T_HISA_Knapp}). We emphasise that also the usage of $T_\rmn{HI, mw}$ -- which is not accessible to an observer -- for $T_\rmn{HISA}$ does not improve the situation but gives qualitatively similar results as using Eq.~\ref{eq:THISATdip}. We attribute this to the fact that $T_\rmn{HI, mw}$ and $T_\rmn{dip}$ are similar within a scatter of 10-20~K (see Fig.~\ref{fig:Tdip_THI}).

To summarise, even when choosing $T_\rmn{HISA}$ by a physically motivated approach, the quality of the obtained HI column density maps does not increase, but partly even decreases. Contrary to the approach of a fixed $T_\rmn{HISA}$, for these approaches not only the quantitative agreement but also the qualitative agreement between $N_\rmn{HI, obs}$ and $N_\rmn{HI, real}$ is lost.

\subsection{The HI velocity dispersion}
\label{sec:dynamics}

\begin{figure*}
\includegraphics[width=\textwidth]{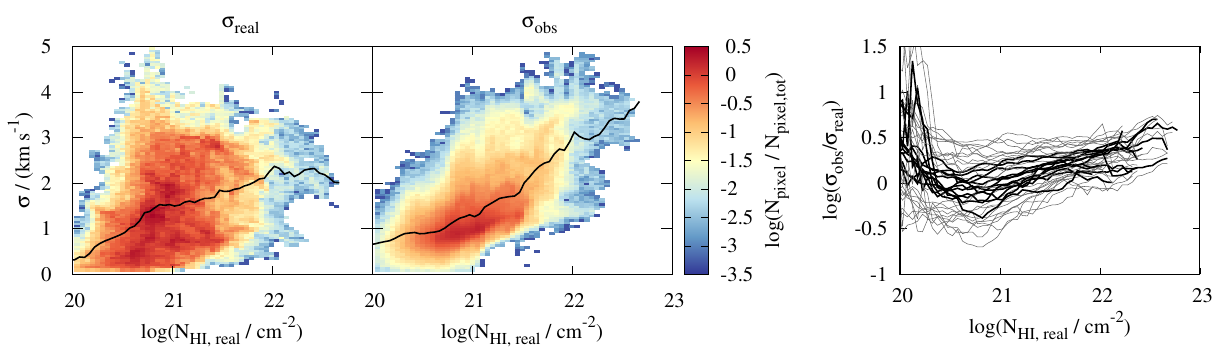} 
\caption{Left and middle panel: Distribution of the real non-thermal HI velocity dispersion, $\sigma_\rmn{real}$, and the observed one, $\sigma_\rmn{obs}$, as a function of $N_\rmn{HI, real}$ for MC1-HD at \mbox{$t_\rmn{evol}$ = 2 Myr} for one direction assuming a distance of 150~pc. Overall, there is an increase of $\sigma$ with the column density although the effect is less pronounced for $\sigma_\rmn{real}$. The stronger increase of $\sigma_\rmn{obs}$ could be due to opacity broadening. Right panel: Ratio of $\sigma_\rmn{obs}$ and $\sigma_\rmn{real}$ as a function of $N_\rmn{HI, real}$. The black line shows the mean for all clouds and three directions at \mbox{$t_\rmn{evol}$ = 2 Myr} assuming a distance of 150~pc, the grey lines represent the interval of one standard deviation. Overall, HISA observations trace the velocity dispersion with an accuracy of a factor of $\sim$2.}
\label{fig:sigma}
\end{figure*}

Finally, we consider the accuracy of the HI velocity dispersion inferred from HISA observations. For this purpose, we compare the non-thermal velocity dispersion $\sigma_\rmn{obs}$ identified via the BTS tool (Section~\ref{sec:HISA_calc}) with the actual HI velocity dispersion along each LOS directly inferred from the simulation data, $\sigma_\rmn{real}$. For $\sigma_\rmn{real}$ we only take into account the velocity component along the LOS and HI with a temperature below 100~K and then calculate for each pixel the HI mass-weighted LOS average. For $\sigma_\rmn{obs}$ we correct the value obtained by BTS, $\sigma_\rmn{BTS}$, for the contribution from the limited channel width of 1~km~s$^{-1}$ and thermal motions, i.e.
\begin{equation}
\sigma_\rmn{obs} = \left( \sigma_\rmn{BTS}^2 - \left( \frac{\textrm{1 km s$^{-1}$}}{\sqrt{8 \, \textrm{ln} \, 2}} \right)^2  - c_\rmn{s}^2 \right)^{1/2} \, ,
\end{equation}
with $c_\rmn{s}$ being the sound speed for HI gas, where, for simplicity, we assume an average temperature of \mbox{40 K} (right panel of Fig.~\ref{fig:Tdip_map}). The factor $\sqrt{8 \, \textrm{ln} \, 2}$ accounts for the conversion of channel width into the standard deviation. In the following we only consider pixels where $\sigma_\rmn{obs}$~$>$~0.

In the left and middle panel of Fig.~\ref{fig:sigma} we plot the distribution of $\sigma_\rmn{real}$ and $\sigma_\rmn{obs}$ and its mean value (black line) as a function of $N_\rmn{HI, real}$ for MC1-HD at \mbox{$t_\rmn{evol}$ = 2 Myr} for one direction assuming a distance of 150~pc for the beam size. We note that the following results also hold for the other clouds and times. First, we find that the scatter for $\sigma_\rmn{real}$ appears to be somewhat larger than for $\sigma_\rmn{obs}$. Second, for $\sigma_\rmn{real}$ there is only a moderate increase with $N_\rmn{HI, real}$ in particular for \mbox{$N_\rmn{HI, real}$ $>$ 10$^{21}$ cm$^{-2}$}, whereas for $\sigma_\rmn{obs}$ the increase is more pronounced.  The latter could be attributed to opacity broadening occurring for high $N_\rmn{HI, real}$ (see black line in Fig.~\ref{fig:spectrum}), which we expect to happen frequently, given the high optical depths found in our MCs (Fig.~\ref{fig:tau}). The presence of multiple Gaussian components in the spectrum (red line in Fig.~\ref{fig:spectrum}), however, can not explain the somewhat larger values of $\sigma_\rmn{obs}$ compared to $\sigma_\rmn{real}$: Although the result of the single-component fit for $\sigma_\rmn{BTS}$ will be broader than the velocity dispersion of the individual HI components, multiple components will also increase $\sigma_\rmn{real}$. We also note that the non-thermal velocity dispersions of HI of a few \mbox{1 km s$^{-1}$} reported here are in general in agreement with the velocity dispersion of dense gas (\mbox{$n$ $>$ 100 cm$^{-3}$}) in these clouds \citep{Seifried17}.

The black lines in the right panel of Fig.~\ref{fig:sigma} show the mean of log($\sigma_\rmn{obs}/\sigma_\rmn{real}$) for all clouds and projection directions at \mbox{2 Myr} and an assumed distance of 150~pc. Despite the differences seen in the left and middle panel, $\sigma_\rmn{obs}$ appears to trace the actual velocity dispersion with a reasonable accuracy. For \mbox{$N_\rmn{HI, real}$ $\lesssim$ 10$^{22}$ cm$^{-2}$}, the mean of log($\sigma_\rmn{obs}/\sigma_\rmn{real}$) is one average within $\pm$0.3~dex around a value of 0, which would indicate a perfect agreement. Also the standard deviations of the various distributions (grey lines) are about \mbox{0.2 -- 0.3 dex}. Hence, we argue that for typical HI column densities between $\sim$10$^{20}$~cm$^{-2}$ and $\sim$10$^{22}$~cm$^{-2}$, HISA observations are able to probe the non-thermal velocity dispersion of HI with an accuracy of a factor of $\sim$2, even in the case of a limited spectral resolution of $\sim$1~km~s$^{-1}$. Only for very high HI column densities (\mbox{$N_\rmn{HI, real}$ $>$ 10$^{22}$ cm$^{-2}$}) the non-thermal velocity dispersion might be somewhat overestimated, which we tentatively attribute to the aforementioned opacity broadening of the absorption feature. We note that these results also hold for \mbox{$t_\rmn{evol}$ = 3 Myr} and an assumed distance of 3~kpc.

\begin{figure}
\includegraphics[width=\linewidth]{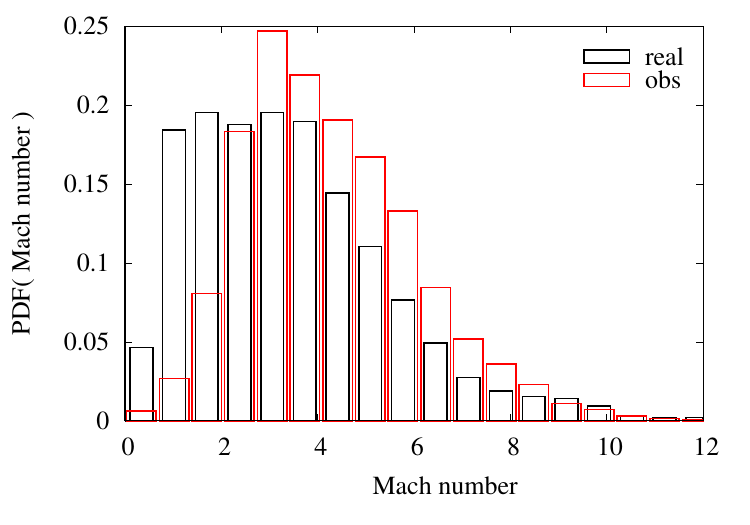} 
\caption{Mach number distribution of the HI gas for the same snapshot as in Fig.~\ref{fig:HI_map} assuming a temperature of 40~K. The black lines shows the distribution obtained from the simulation data directly ($\sigma_\rmn{real}$), the red that from the observed velocity dispersion ($\sigma_\rmn{obs}$). The HI gas is moderately supersonic with Mach numbers around 1 -- 10.}
\label{fig:mach}
\end{figure}
Finally, in Fig.~\ref{fig:mach} we show the distribution of the Mach numbers $\sqrt{3}\sigma/c_\rmn{s}$ (assuming \mbox{$T$ = 40 K}) for MC1-HD at \mbox{2 Myr}. The values are inferred directly from the simulation using $\sigma_\rmn{real}$ (black lines) and from the HISA observation using $\sigma_\rmn{obs}$ (red lines). For other clouds, we find similar results. Overall, the HI gas is moderately supersonic with the distribution peaking around Mach numbers of a few although also values up to $\sim$10 are reached. The observationally determined Mach numbers somewhat underrepresent the lowest values. This could be related to the aforementioned opacity broadening or a broadening due to the limited spectral resolution. The overall similarity between both distributions, however, confirms the accuracy of a factor of $\sim$2 between $\sigma_\rmn{real}$ and $\sigma_\rmn{obs}$ reported before. Furthermore, the Mach numbers found are also in good agreement with recent HISA observations \citep{Burkhart15,Nguyen19,Syed20,Wang20}.

\section{Discussion}
\label{sec:discussion}

\subsection{The $N_\rmn{HI}$-PDF: Comparison with observations}
\label{sec:obs}

\begin{figure}
\includegraphics[width=\linewidth]{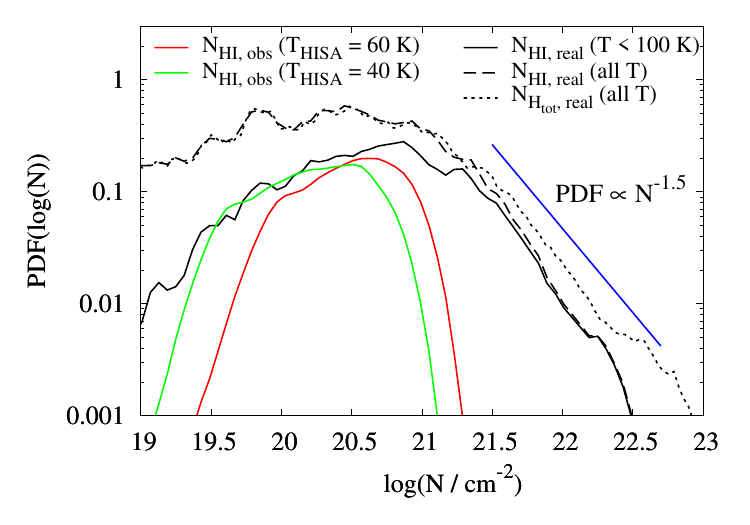} 
\caption{Column density PDFs of the cold HI (black solid line), the total HI (black dashed line), and all hydrogen nuclei (H$_\rmn{tot}$, black dotted line) inferred directly from the simulation of MC1-HD at 2~Myr for one direction. In addition, the corresponding PDFs determined from the synthetic HISA observations assuming \mbox{$T_\rmn{HISA}$ = 40 and 60 K} (green and red line, respectively) are shown. The $N_\rmn{HI, obs}$-PDFs peak at significantly lower values than that of $N_\rmn{HI, real}$ and show a roughly lognormal distribution. In contrast, the $N_\rmn{HI, real}$($N_\rmn{H, tot}$)-PDFs exhibit a power-law tail roughly proportional to $N^{-1.5}$ (blue line), which indicates that also  HI is gravitationally unstable.}
\label{fig:PDFs}
\end{figure}

As discussed in Section~\ref{sec:T_HISA_fixed}, we find that HISA measurements tend to underestimate the actual column density of cold HI by a factor of 3 -- 10 in the outer parts of MCs and potentially by an even higher factor in the central, high density parts. We attribute this to (i) a large temperature variation of the HI gas and the assumption of a fixed $T_\rmn{HISA}$ and (ii) the effect of noise in the brightness temperature measurements in particular for regions of high optical depth.

This underestimation is again demonstrated in Fig.~\ref{fig:PDFs} where we show the area-weighted PDFs of $N_\rmn{HI, real}$ and $N_\rmn{HI, obs}$ for MC1-HD at \mbox{$t_\rmn{evol}$ = 2 Myr} for one direction at an assumed distance of 150~pc\footnote{We note that the integrated area under the curves are not necessarily unity as many pixels have no observed HI column density (Fig.~\ref{fig:HI_map}). Hence, the relative area under the curves gives the reader a direct indication of how many pixels are omitted (i.e. have \mbox{$N_\rmn{HI, obs}$ = 0}) compared to the PDF of all HI.}. The latter is derived for \mbox{$T_\rmn{HISA}$ = 40 and 60 K} (coloured lines). As expected, the peak of the $N_\rmn{HI, obs}$-PDF is shifted by a factor of 3 -- 10 towards lower column densities with respect to that of the $N_\rmn{HI, real}$-PDF (black solid line). Moreover, also the shapes of the two PDFs are very different with important implications. First, the $N_\rmn{HI, real}$-PDF is significantly broader than the $N_\rmn{HI, obs}$-PDF. This indicates that the width of the $N_\rmn{HI, obs}$-PDF obtained from HISA observations might not be a good quantity to assess turbulence statistics \citep[see e.g.][for an application to the $N_\rmn{H, tot}$-PDF]{Burkhart12}. Second, the $N_\rmn{HI, real}$-PDF shows signs of a power-law tail at column densities above a few 10$^{21}$~cm$^{-2}$, which is roughly proportional to $N^{-1.5}$, similar to that of the $N_\rmn{H, tot}$-PDF \citep[black dotted line, see also][]{Kainulainen09,Kritsuk11,Girichidis14,Schneider15,Auddy18,Veltchev19}. This power-law tail indicates that also the dense HI gas is undergoing gravitational collapse. We emphasise, however, that -- as the dense gas (\mbox{$N \gtrsim 10^{21}$ cm$^{-2}$}) is predominantly molecular (bottom panel of Fig.~\ref{fig:NH-diag}) -- the gravitational force in this range is dominated by gas in form of H$_2$ with which the HI is mixed.

Our $N_\rmn{HI, real}$-PDF are markedly different from PDFs found in recent HISA observations \citep{Burkhart15,Imara16,Syed20,Wang20} which find a log-normal shape indicating that the cold HI is not gravitationally unstable. This apparent contradiction could have its origin in possible observational biases, indicated by some striking similarities between our synthetic HI observations and that of the aforementioned authors: Our synthetic and the actually observed $N_\rmn{HI, obs}$-PDF are of roughly lognormal shape, are in a similar range (\mbox{$N_\rmn{HI}$ = 10$^{20 - 21}$ cm$^{-2}$}), and are at significantly lower column densities than that of either H$_2$ and the total HI observed in emission \citep{Syed20,Wang20} or that of $N_\rmn{HI, tot}$ measured via dust emission \citep[][]{Burkhart15,Imara16}. Taking these similarities into account, we suggest that there indeed is an observation bias in the shape of observed $N_\rmn{HI}$-PDFs. This could be particularly pronounced at the high end of the PDF, which is often characterised by a power-law. In addition, the $N_\rmn{HI}$ values obtained should rather be considered as lower thresholds. We note, however, that an MC at an very early evolutionary stage might not have undergone gravitational collapse, i.e. might not yet have developed high column densities (\mbox{$\gtrsim$ 10$^{21}$ cm$^{-2}$}) and the associated power-law tail in the $N$-PDF. Therefore, for such an MC, the assessment of the HI column densities and masses via HISA observations might still be somewhat better compared to the findings presented here.

\subsection{Multiple HI-H$_2$ transitions: Comparison with analytical results}
\label{sec:layers}

As stated before, recent semi-analytical works predict that, for ISM conditions comparable to that of the solar neighbourhood, the transition from HI to H$_2$ occurs at column densities of \mbox{$\lesssim 10^{21}$ cm$^{-2}$} \citep{Krumholz08,Krumholz09,Sternberg14,Bialy16}. These models also suggest an upper limit of $N_\rmn{HI}$ around this value. Contrary to that, for the clouds simulated in this work, we find HI column densities partly well above this value (see Figs.~\ref{fig:NH-diag} and~\ref{fig:HI_emission}). Also recent observations of W43 \citep{Motte14,Bihr15} and indirect estimates towards Perseus \citep{Okamoto17} and clouds outside the Galactic plane \citep{Fukui14,Fukui15} have revealed HI column densities of up to a few \mbox{10$^{22}$ cm$^{-2}$}, thus well comparable to our findings, but in apparent contradiction to the theoretical predictions. Also observations of the Magellanic Clouds by \citet{Welty12} seem to challenge the prediction for the value of $N_\rmn{H, tot}$, where the transition to H$_2$ dominated gas is supposed to occur \citep{Krumholz08,Krumholz09,McKee10}.

However, strictly speaking the suggested, upper limit around 10$^{21}$~cm$^{-2}$ only applies to a single HI-H$_2$ transition. As pointed out by \citet{Motte14}, a possible solution for this contradiction could thus be the presence of several transitions along the LOS \citep[see also][]{Bialy17}. Indeed, the occurrence of multiple absorption components (see Fig.~\ref{fig:spectrum}) and the highly complex structure of our simulated MCs (see Fig.~\ref{fig:HI_emission}) and of real MCs indicates that the assumption of a single HI-H$_2$ transition might be an oversimplification and that rather several transitions are present.

\begin{figure}
\includegraphics[width=\linewidth]{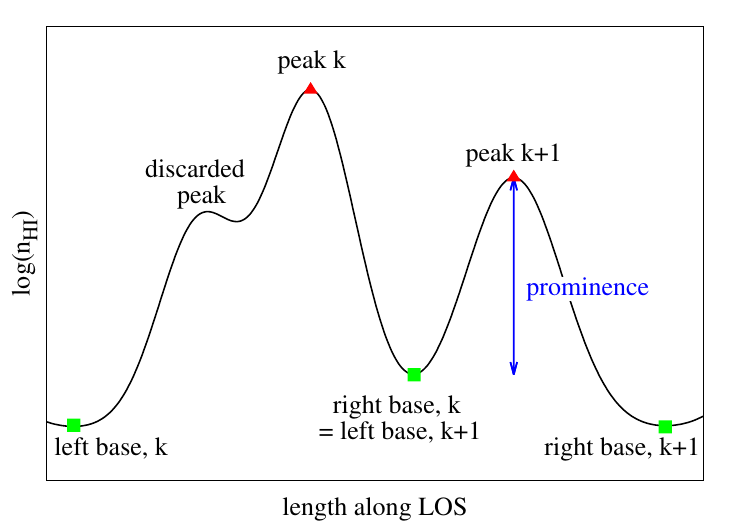} 
\caption{Sketch for the identification of HI peaks (red triangles) and the position of their bases (green squares), also for a case where a local peak is discarded. The blue arrow indicates the prominence of peak k+1.}
\label{fig:peak-sketch}
\end{figure}
In the following we test this hypothesis in physical space (i.e. considering spatial distances) as opposed to parts of the analysis done before in velocity space (Section~\ref{sec:results}). For this purpose, we identify the number of HI-density peaks along rays which intersect the entire length of the zoom-in regions and which are distributed uniformly over the entire area of the emission maps with a spacing of 0.24~pc. For this purpose, we first determine the profile of the logarithm of the HI density, log($n_\rmn{HI})(l)$, along each ray, i.e. along the LOS of each pixel. A few selected profiles are shown in Fig.~\ref{fig:profiles}, demonstrating their variability for different pixels, which necessitates a more systematic study. Second, we identify the positions of the peaks of log($n_\rmn{HI})$ along the LOS, $l_\rmn{peak, k}$, as well as the position of their left and right base, $l_\rmn{base,left, k}$ and $l_\rmn{base, right, k}$, respectively (see Fig.~\ref{fig:peak-sketch} for a schematic view). The base of a peak (green squares) is the minimum of the density profile between this and the neighbouring peak. In order to account for the influence of small-scale density fluctuations, we discard peaks which (i) have a prominence (blue arrow) of less then 0.5 (in log-space), or (ii) have a peak height log($n_\rmn{HI, peak}$) of less than 0.5 ($n_\rmn{HI, peak}$~$\simeq$~3~cm$^{-3}$), or (iii) are separated from the next (and higher) peak by less then 2~grid cells along the LOS\footnote{When we discard a peak, the minimum between this discarded and the neighbouring peak $k$, is \textit{not} taken as the base of the peak $k$. Rather, the base of the peak $k$ is shifted beyond the discarded peak such that the discarded peak now lies between $l_\rmn{base, k}$ and $l_\rmn{peak, k}$ (see left-most green square in Fig.~\ref{fig:peak-sketch}). We also note that we tested the approach by discarding the lower of two peaks when they are separated by less than 4 cells. This, however, affected the findings only marginally, which is why we do not follow this further here.}. Furthermore, we note that as $n_\rmn{HI}$ comes from the chemical network implemented in the simulations, a low $n_\rmn{HI}$ could indicated either a low total gas density or a high gas density, where hydrogen is already predominantly in form of H$_2$.

\begin{figure*}
\includegraphics[width=\linewidth]{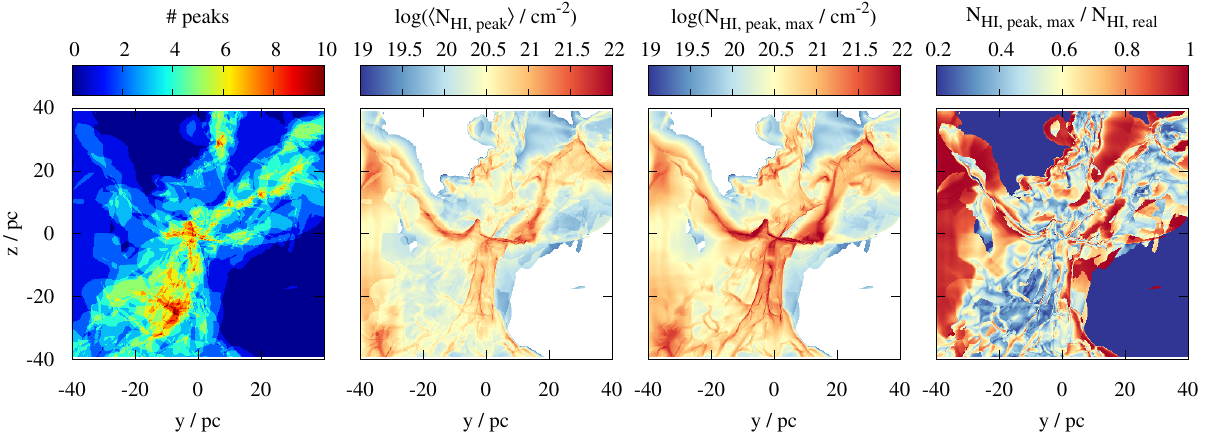} 
\caption{Maps of the number of HI-density peaks per LOS, the mean HI column density per HI-density peak, the HI column density of the most prominent HI-density peak and the ratio of this value to the total HI column density (from left to right). The left-most panel demonstrates the occurrence of up to 10 peaks, which corresponds to the same number of HI-H$_2$ transitions along the LOS. In consequence, the average HI column density per peak (center-left panel) mostly matches the theoretical predictions of \mbox{$\lesssim 10^{21}$ cm$^{-2}$}. However, the most prominent HI-density peak (center-right panel) often has a column density that is considerably larger than this value and contributes significantly to $N_\rmn{HI, real}$ ($\gtrsim$~40\%, right panel).}
\label{fig:peaks}
\end{figure*}
Next, we calculate the column density of each HI-density peak as
\begin{equation}
 N_\rmn{HI, peak, k} = \int_{l_\rmn{base,left, k}}^{l_\rmn{base,right, k}} n_\rmn{HI} \rmn{d}l \, ,
\end{equation}
and determine the average column density, $\langle N_\rmn{HI, peak} \rangle$, of all peaks along a given LOS. In addition, we identify the most massive peak, i.e. the peak which accumulates most of the HI along the given LOS, and determine its column density, $N_\rmn{HI, peak,max}$. In Fig.~\ref{fig:peaks}, we plot the number of peaks along each LOS, $\langle N_\rmn{HI, peak} \rangle$, $N_\rmn{HI, peak, max}$ and the ratio $N_\rmn{HI, peak, max}$/$N_\rmn{HI, real}$ for MC1-HD at 2 Myr along one direction. The last value shows how much the most massive peak contributes to the overall HI column density.

The left panel of Fig.~\ref{fig:peaks} shows that there are indeed up to $\sim$10 HI-density peaks along the LOS as suggested by \citet{Bialy17} for the case of W43. There is a moderate tendency of a higher number of peaks towards the center of the MC, i.e. with increasing $N_\rmn{HI, real}$, probably caused by the filamentary substructure of the MCs. This increase causes $\langle N_\rmn{HI, peak} \rangle$ to remain below 10$^{21}$~cm$^{-2}$ for the vast majority of rays (second panel of the left), only for about 5 -- 25\% of all rays (depending on the cloud, direction, and time) it exceeds this value. Hence, on first view this appears to be in rough agreement with analytical predictions \citep{Krumholz08,Krumholz09,Sternberg14,Bialy16}.

However, the column density of the dominant peak, $N_\rmn{HI, peak, max}$, exceeds the value of 10$^{21}$~cm$^{-2}$ for a large number of rays (second panel from the right). We find that $N_\rmn{HI, peak, max}$ exceeds 10$^{21}$~cm$^{-2}$ for 30 -- 50\% of the rays, i.e. more than twice as often as the corresponding fraction for $\langle N_\rmn{HI, peak} \rangle$. Again, the exact fraction depends on the considered MC, direction, and time. However, we do not see any dependence on the presence or absence of magnetic fields in the simulations, despite the fact that the field is dynamically important for the overall (chemical) evolution of the MCs \citep[see][]{Seifried20,Seifried20b}. Furthermore, the dominant peak accounts on average for about 40 -- 60\% and even more of $N_\rmn{HI, real}$ (right panel) and is thus indeed dominating the overall HI budget of the clouds. Hence, these findings are in contrast to the theoretical models, and we will discuss their implications in detail in the following section.

\section{Is there more cold HI than thought?}
\label{sec:moreHI}

\subsection{The theoretical perspective}

As discussed so far, our results indicate that semi-analytical models tend to underestimate the maximum column density of cold HI in MCs. This can be attributed to several reasons. First, as suggested by \citet{Motte14} and \citet{Bialy17} and explicitly shown here for the first time, in realistic models of MCs there appear up to $\sim$10 HI-H$_2$ transitions along the LOS. Secondly, non-equilibrium effects can increase the HI content of MCs due to the limited time available for H$_2$ to form \citep{Glover07b,Glover10,MacLow12,Motte14}. We investigate this effect by artificially evolving the chemistry for a selected snapshot to chemical equilibrium (see Appendix~\ref{sec:appendixa} for details). Doing so, we find that this reduces the HI content in our MCs by a factor of 2 -- 2.5 (see Fig.~\ref{fig:HI_equilibrium}). Hence, the assumption of chemical equilibrium in semi-analytical models indeed results in too low HI abundances compared to the actual non-equilibrium HI present in dynamically evolving MCs.

\begin{figure*}
\includegraphics[height=0.32\textwidth]{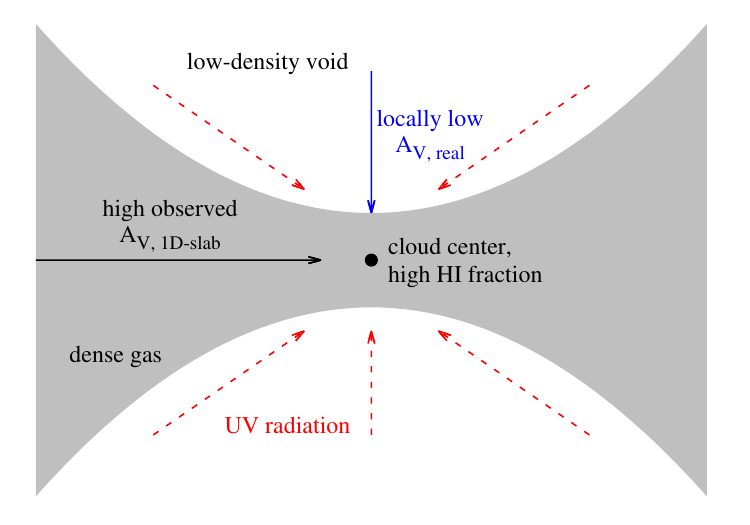}\hspace{2cm}
\includegraphics[height=0.32\textwidth]{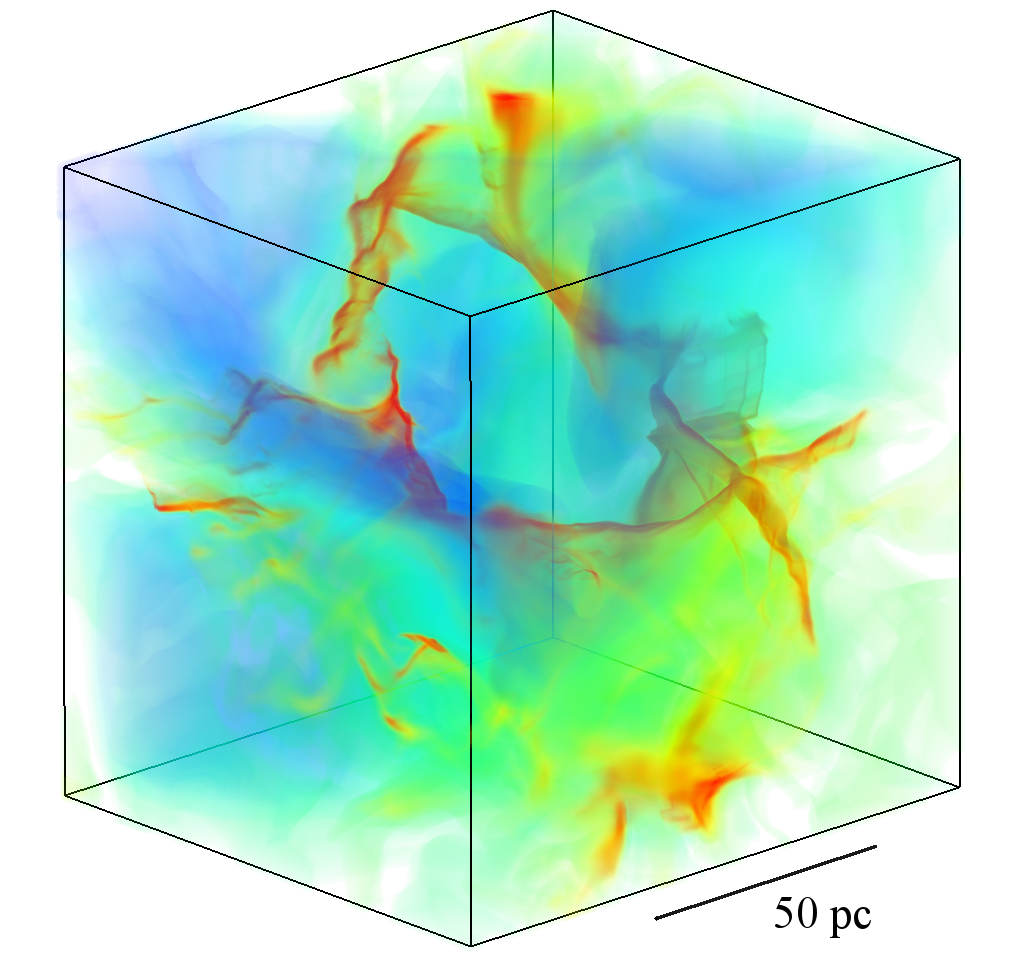}
\caption{Left: Sketch of a situation in a MC (dense gas depicted with grey). Here, the observer's assumption of a 1-dimensional, plane-parallel configuration as applied in semi-analytical models overestimates the shielding (black arrow, $A_\rmn{V, 1D-slab}$) and thus underestimates the amount of HI in the cloud's center. Under realistic conditions, radiation (red arrows) dissociating H$_2$ might be able to propagate through low-density voids (white areas) towards the center of the cloud. This results in a rather low actual visual extinction, $A_\rmn{V, real}$. Right: Volume rendering of MC1-HD at 2~Myr showing the highly complex and filamentary structure of the dense gas (reddish). Large low-density voids (bluish) are recognisable through which radiation can propagate into the cloud thus increasing the HI fraction. This resembles the simplified picture shown in the left panel.}
\label{fig:AV}
\end{figure*}
In addition, we here suggest a third reason why even for a single HI-H$_2$ transition $N_\rmn{HI}$ could be higher (center-right panel of Fig.~\ref{fig:peaks}) than predicted by semi-analytical models with a 1-dimensional geometry \citep{Krumholz08,Krumholz09,Sternberg14,Bialy16}. For this purpose we consider the effect of assuming an idealised plane-parallel slab in more detail. In such a configuration the gas in the cloud is irradiated only from one side, i.e. a parcel of gas having a high column density in the slab direction will receive very little radiation. In other words, for a plane-parallel slab -- at a given total hydrogen nuclei column density $N_\rmn{H, tot}$ towards the direction of the incident radiation -- the extinction is maximal as radiation coming from other possible directions is neglected. In consequence, also the amount of HI is minimal due to the lack of H$_2$ dissociation. The same chain of arguments can also be made for a spherically symmetric configuration.

In contrast, under realistic conditions in turbulent MCs, there may exist large low-density voids through which the radiation can propagate into the cloud (almost) unhindered. This is sketched in the left panel of Fig.~\ref{fig:AV}: a dense region shielded completely against UV radiation from one direction (horizontal) can still be irradiated from another direction (vertical). Hence, if observed from the horizontal direction, this region would \textit{appear} completely optically thick (i.e. a high value of $A_\rmn{V, 1D-slab}$), corresponding to the assumption of a plane-parallel configuration. In reality, however, the region might still receive up to $\sim$50\% of the radiation coming from the vertical direction, i.e. the real $A_V$ might be significantly lower\footnote{In order to calculate the average visual extinction in a point, one must not take the average $\langle A_\rmn{V,i} \rangle$ over different directions i, but must take the logarithm of $\langle \rmn{exp}(-\gamma A_\rmn{V,i} \rangle$, as the latter describes the amount of incident radiation. The latter averaging puts more emphasis on the low-$A_\rmn{V}$ directions.}. In consequence, H$_2$ might still be dissociated by UV radiation even in the central regions of MCs thus increasing the amount of HI. Considering a 3-dimensional graphical representation of MC1-HD at 2~Myr (right panel of Fig.~\ref{fig:AV}) indeed shows that there exist these large low-density voids (bluish) through which radiation can travel (almost) unattenuated thereby dissociating H$_2$ in the densest regions (reddish). Depending on the considered snapshot, 55 to 83\% of the volume of the zoom-in regions is covered by gas with densities below 1~cm$^{-3}$. We note that also for internal, stellar radiative feedback the actual cloud's substructure and shielding parameters have a similar importance \citep{Haid19}.

One could interpret this effect as either increasing the radiation strength or decreasing the \textit{effective} shielding of UV radiation in the cloud. In the semi-analytical models of \citet{Sternberg14}, \citet{Bialy16} and \citet{Bialy17b}, this is parametrised by their parameter $\alpha G$. There, $\alpha$ indicates the radiation strength and $G$ the shielding factor including H$_2$ self-shielding and dust attenuation; the lower $G$, the better the radiation is shielded. We speculate that large density voids can easily reduce the (self-)shielding of the surrounding dust and H$_2$ by an order of magnitude. The corresponding \textit{increase} of $G$ (and thus $\alpha G$) in the models of the aforementioned authors would thus increase their $N_\rmn{HI}$ \citep[see e.g. equation~40 and figure~9 in][]{Sternberg14} by a factor of a few. This behaviour is thus in general accordance with the interpretation presented here and would bring their upper limits for $N_\rmn{HI}$ closer to the values reported here.

\subsection{The observational perspective}

As observational works on Galactic \citep[e.g.][]{Savage77,Kavars03,Kavars05,Klaassen05,Gillmon06,Krco08,Barriault10,Krco10,Lee12,Lee15,Stanimirovic14,Burkhart15,Imara16} and extragalactic scales \citep[e.g.][]{Wong02,Browning03,Blitz04,Blitz06,Bigiel08,Wong09} tend to find upper limits of $N_\rmn{HI}$ around 10$^{21}$~cm$^{-2}$ (equivalent to \mbox{8 M$_{\sun}$ pc$^{-2}$}), it could be argued that the high HI column densities in MCs and the associated underestimation of HI by a factor of 3-10 suggested in this work are rather exceptional. Given the various reasons discussed in this work, however, we argue that the underestimation is indeed rather common. In addition, carefully investigating the observational literature we find further evidence that an upper $N_\rmn{HI}$ threshold of 10$^{21}$~cm$^{-2}$ could be artificial:
\begin{itemize}
\item Some of the aforementioned HI measurements are emission observations, and for some of them the contribution of the cold HI might be omitted, e.g. \citet{Lee12} report two absorption features in their HI spectra of the Perseus molecular clouds, which they do not consider in their $N_\rmn{HI}$ calculations. Furthermore, other highly-resolved HI emission observations indeed find HI column densities well above 10$^{21}$~cm$^{-2}$ \citep{Motte14,Bihr15,Syed20,Wang20}.
\item Observations in emission often assume optically thin emission. However, as large parts of MCs have optical depths well above 1 (Fig.~\ref{fig:tau}), optical depth corrections are crucial to infer the correct HI column densities. This is in line with observations of W43 by \citet{Bihr15}, who find an increase in HI by a factor of $\sim$2 compared to the optically thin assumption \citep{Motte14}. Similar correction factors were found for indirect HI measurements of off-Galactic plane gas \citep{Fukui15} and the Perseus molecular cloud \citep{Okamoto17}, although for the latter \citet{Lee12,Lee15} argue for a correction of $\sim$20\% only. For the THOR survey \citet{Wang20b} determined the correction factor to $\sim$31\%. All these correction factors are lower limit as optical depth estimates have an upper limit set by the observational noise \citep[][and our Eq.~\ref{eq:deltatau}]{Bihr15}.
\item Finally, extragalactic observations typically have spatial resolutions of a few 100~pc. Hence, they average over clouds and the surrounding diffuse ISM, which can lower the maximum value of $N_\rmn{HI}$ significantly. We show this by calculating the total HI column density (now including again HI with \mbox{$T$ $>$ 100 K}) for our simulations using pixels with a side length of 31.5, 8 and 2~pc (black, magenta and green dots, respectively, in the top panel of Fig.~\ref{fig:NH-diag}). Overall, this reduces the maximum $N_\rmn{HI}$ values to $\sim$10$^{22}$~cm$^{-2}$ for 2~pc pixels and even further to  $\sim$10$^{21}$~cm$^{-2}$ for 31.5~pc pixels, i.e. in parts by more than one order of magnitude compared to the maximum around a few 10$^{22}$~cm$^{-2}$ for our highest resolution.
\end{itemize}

To summarize, we suggest that (i)~HI column densities well beyond 10$^{21}$~cm$^{-2}$ ($\sim$8~M$_{\sun}$~pc$^{-2}$) are significantly more common in MCs than thought (Fig.~\ref{fig:N_real_vs_obs}) and (ii)~also the entire mass of cold HI gas in clouds could be a factor of a few higher ($\gtrsim 3$) than thought (Fig.~\ref{fig:masses}). Vice versa, we argue that (iii)~1-dimensional PDR models might underestimate the amount of cold HI in a typical HI-H$_2$ transition layer as those are in general not plane-parallel or spherically symmetric objects and (iv)~HI observations might underestimate the HI content in MCs by a factor of a few due to the various systematic observational biases discussed in this work.

\section{Conclusions}
\label{sec:conclusion}

In this work we present the first fully self-consistent synthetic HI~21~cm observations including self-absorption (HISA) of MCs simulated within the SILCC-Zoom project. The synthetic observations are based on 3D MHD simulations including a non-equilibrium HI-H$_2$ chemistry, detailed radiative transfer calculations, and realistic observational effects like noise and a limited spectral and spatial resolution adapted to actual observations. In addition, we analyse in detail the actual content of cold HI in the simulated clouds and compare it with the results obtained from the synthetic HISA observations. We summarize our main results in the following.

\begin{itemize}
\item We show that HISA observations, which assume a fixed HI temperature, typically tend to underestimate column densities of cold HI, $N_\rmn{HI}$, and the total cold HI mass in molecular clouds by a factor of 3 -- 10. This effect is particular pronounced towards to the central regions, which frequently reach column densities up to $\gtrsim$10$^{22}$~cm$^{-2}$. It occurs for MCs under various conditions, e.g. with and without dynamically important magnetic fields.
\item We show that the underestimation of $N_\rmn{HI}$ in HISA observations can be attributed to the following two effects. (i) The large temperature variations of cold HI ($\sim$10~K up to 100~K) make a reliable determination of $T_\rmn{HISA}$ not possible. This leads to the fact that the real  $T_\rmn{HISA}$ and thus $N_\rmn{HI}$ are underestimated and that velocity channels have to be omitted for the calculation of $N_\rmn{HI}$. (ii) Observational noise and the emission of warm HI in the foreground either reduce the inferred optical depth or  -- as before -- cause individual velocity channels to be omitted for the calculation of $N_\rmn{HI}$. This effect is particularly pronounced in regions of high optical depth. In combination, both effects (i + ii) can lead to in an artificial upper limit in observation of $N_\rmn{HI, obs}$ around 10$^{21}$~cm$^{-2}$.
\item We suggest a method to correct for the aforementioned omission of high optical depth channels. This correction reduces underestimation of the HI mass budget by a factor of 1.5 -- 2.
\item We find that clouds typically have HI optical depths around 1 -- 10. This implies that the optically thin HI assumption is usually not suitable and that optical depth corrections are essential when calculating $N_\rmn{HI}$ from HI observations.
\item We show that the high HI column densities ($\gtrsim10^{22}$~cm$^{-2}$) can (in parts) be attributed to the occurrence of up to 10 individual HI-H$_2$ transitions along the LOS. This emphasises the necessity of Gaussian decomposition algorithms to fully analyse the individual components that constitute the HISA spectra.
\item Also for individual HI-H$_2$ transitions, $N_\rmn{HI}$ frequently exceeds a value of 10$^{21}$~cm$^{-2}$, thus challenging 1-dimensional, semi-analytical models. This can be attributed to non-equilibrium chemistry effects, which are included in our models, and to the fact that HI-H$_2$ transitions usually do not have a 1-dimensional geometry, i.e. to the fractal structure of MCs. 
\item We demonstrate that $N_\rmn{HI}$-PDFs obtained from HISA observations with a fixed temperature assumption should be considered with great caution both concerning the position of the peak and the width. Due to the underestimation of HI, the observed PDFs appear to lack the high-$N_\rmn{HI}$ end, which in reality seems to be characterised by a power-law.
\item Finally, we show that the cold HI gas in MCs is moderately supersonic with Mach numbers of up to a few. The corresponding non-thermal velocity dispersion can be determined via HISA observations with an accuracy of a factor of $\sim$2.
\end{itemize}

To summarize, our result indicate that measuring the HI content in MCs via HISA observations is a challenging task and that the amount of cold HI in MCs could be a factor of 3 -- 10 higher than previously thought.

\section*{Acknowledgements}

We thank the referee for a very thorough and helpful report, which helped to improved the clarity of this work. DS likes to thank H. D\'{e}nes for helpful discussions. DS and SW acknowledge support of the Bonn-Cologne Graduate School, which is funded through the German Excellence Initiative as well as funding by the Deutsche Forschungsgemeinschaft (DFG) via the Collaborative Research Center SFB 956 ``Conditions and Impact of Star Formation'' (subprojects C5 and C6). SW acknowledges support via the ERC starting grant No. 679852 "RADFEEDBACK". HB and JDS acknowledge support from the European Research Council under the Horizon 2020 Framework Program via the ERC Consolidator Grant CSF-648505.  HB and JS acknowledge support from the DFG in the Collaborative Research Center SFB 881 - Project-ID 138713538 - ``The Milky Way System'' (subproject B1). PG acknowledges funding from the European Research Council under ERC-CoG grant CRAGSMAN-646955 and the ERC Synergy Grant ECOGAL (grant 855130). RW aknowledges support by project 19-15008S of the Czech Science Foundation and by the institutional project RVO:67985815. The FLASH code used in this work was partly developed by the Flash Center for Computational Science at the University of Chicago. The authors acknowledge the Leibniz-Rechenzentrum Garching for providing computing time on SuperMUC via the project ``pr94du'' as well as the Gauss Centre for Supercomputing e.V. (www.gauss-centre.eu).

\section*{Data Availability}

The data underlying this article can be shared for selected scientific purposes after request to the corresponding author.




\bibliographystyle{mnras}
\bibliography{literature} 



\appendix

\section{Supplementary figures}
\label{sec:appendixa}

\begin{figure}
\centering
\includegraphics[width=\linewidth]{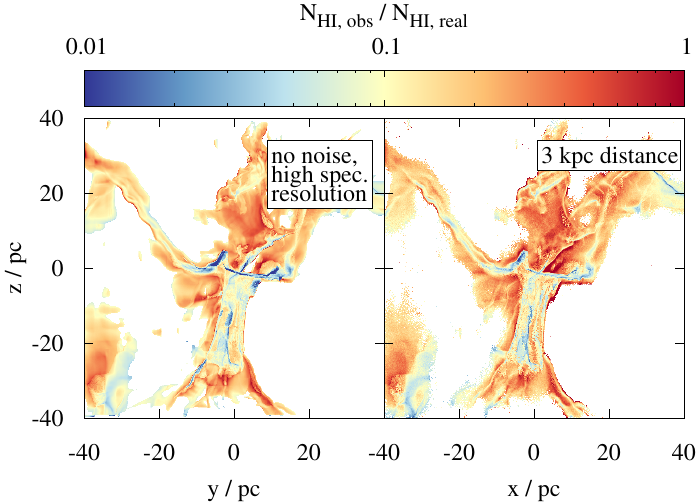} 
\caption{Map of the ratio of the observed and actual HI column density for the same snapshot as in Fig.~\ref{fig:HI_map}, now for the case of no noise and a high spectral resolution (left) and an assumed distance of 3~kpc including observational effects like noise (right) inferred from the HISA observation assuming \mbox{$T_\rmn{HISA}$  = 40 K}. The poor match is found also for the ideal observations (left) supporting the conclusion that strong temperature variation are (in parts) the cause for it.}
\label{fig:HI_map_no_noise}
\end{figure}

In the left panel of Fig.~\ref{fig:HI_map_no_noise} we show ratio of $N_\rmn{HI, obs}$ to $N_\rmn{HI, real}$ for MC1-HD at 2 Myr, where $N_\rmn{HI, obs}$ is calculated from the noiseless, high-spectral resolution (\mbox{200 m s$^{-1}$}) maps obtained directly from RADMC-3D using \mbox{$T_\rmn{HISA}$ = 40 K}. We find a comparable poor match as for the case when observational effects are included. In the right panel we show the results obtained assuming a distance of 3~kpc (again including observational effects). Little differences are found compared to a distance of 150~pc (compare with the bottom middle panel of Fig.~\ref{fig:HI_map}). This result holds also for the other snapshots considered in this work.

\begin{figure*}
\centering
\includegraphics[width=0.9\linewidth]{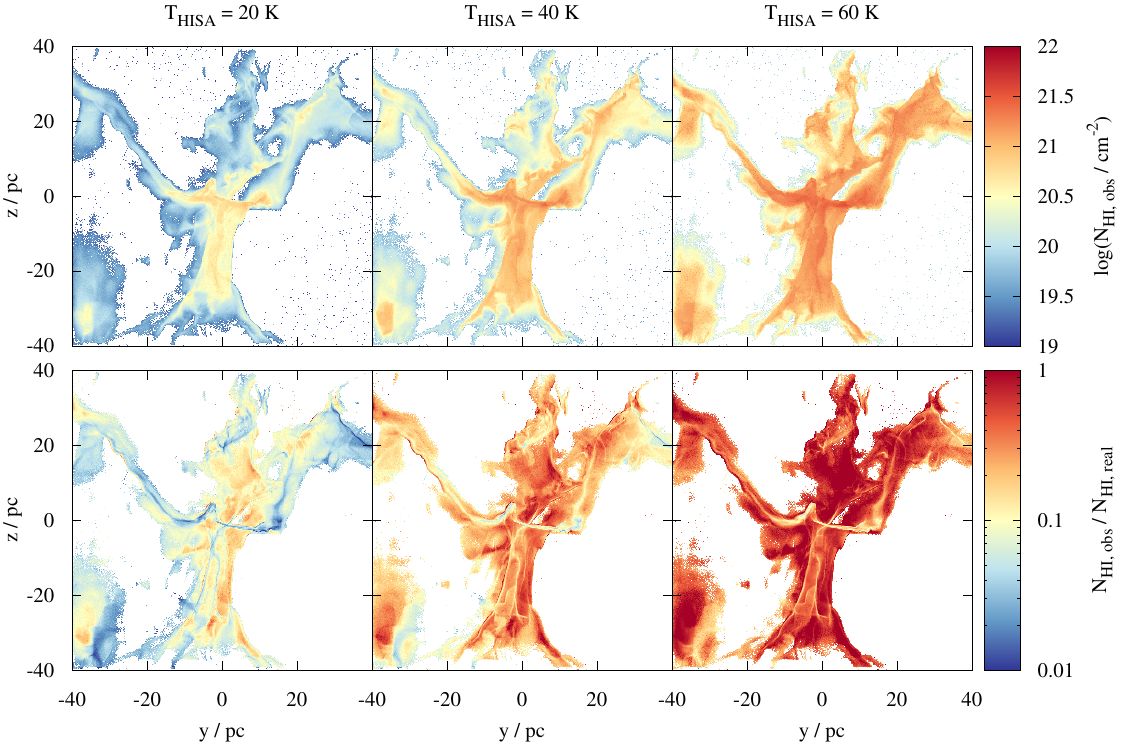} 
\caption{Same as in Fig.~\ref{fig:HI_map}, but now with the correction in optically thick regions where for channels, where Eq.~\ref{eq:tau_HISA} does not yield any result, we assume an optical depth given by $\tau_\rmn{HI, noise}$ (Eq.~\ref{eq:deltatau}). Overall, the match in the moderately dense gas is improved, whereas in the most densest parts $N_\rmn{HI, real}$ is still significantly underestimated.}
\label{fig:HI_map_added}
\end{figure*}
In Fig.~\ref{fig:HI_map_added} we show the inferred HI column density for MC1-HD at 2 Myr now including the correction in optically thick regions. The results are discussed in Section~\ref{sec:opacity_correction}.

\begin{figure}
\centering
\includegraphics[width=0.9\linewidth]{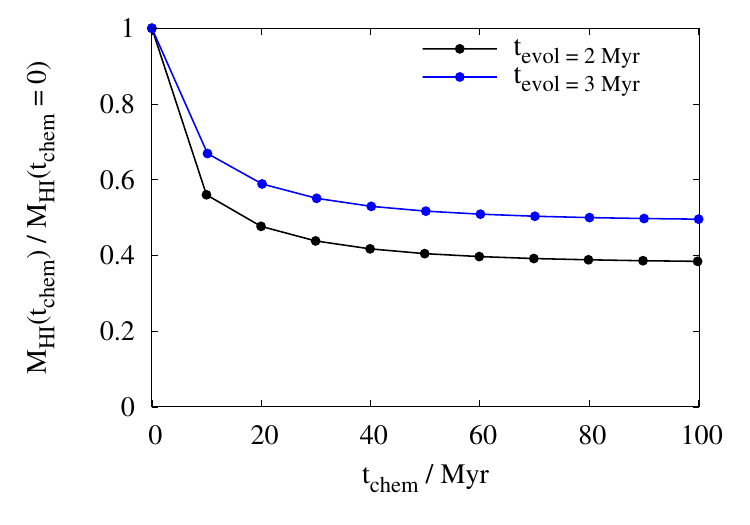} 
\caption{Evolution of the HI content of MC1-HD when post-processing the chemical abundances for a time of $t_\rmn{chem}$ relative to the actual (non-equilibrium) HI content (at \mbox{$t_\rmn{chem}$ = 0}). Assuming chemical equilibrium would reduce the HI content by a factor of 2 -- 2.5.}
\label{fig:HI_equilibrium}
\end{figure}
In Fig.~\ref{fig:HI_equilibrium} we show the effect of post-processing the chemical state of one of our simulation, i.e. pushing it towards a chemical equilibrium state. This is done exemplarily for MC1-HD taking two snapshots at \mbox{$t_\rmn{evol}$ = 2} and 3~Myr considered in this work. We stop the magneto-hydrodynamical evolution at these time, i.e. freeze the total density, velocity, etc., and only evolve the chemistry for additional 100 Myr. The chemical post-processing time (measured from $t_\rmn{evol}$ onwards) is denoted as $t_\rmn{chem}$. The HI content quickly drops to 40 -- 50\% of the the actual (non-equilibrium) HI content (at \mbox{$t_\rmn{chem}$ = 0}), which indicates that equilibrium models generally underestimate the the amount of HI in MCs.

\begin{figure}
\centering
\includegraphics[width=0.9\linewidth]{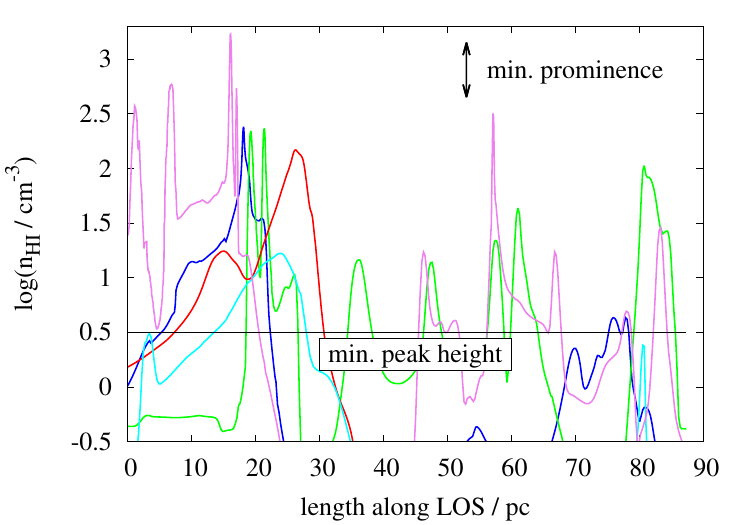} 
\caption{Profiles of log($n_\rmn{HI}$) for 5 selected pixels for MC1 at 2~Myr along the $x$-direction. The profiles show a large variability. In addition we show the minimum prominence and minimum peak value (both in black), which a peak must have to be considered and not discarded.}
\label{fig:profiles}
\end{figure}
In Fig.~\ref{fig:profiles} we show the log($n_\rmn{HI}$)-profile for 5 selected pixels for MC1 at 2~Myr along the $x$-direction used for the analysis in Section~\ref{sec:layers}. The profiles show a large variability concerning the number of peaks, their widths and positions. Some of the peaks are discarded as they do not have either the required minimum prominence or the minimum peak height.

\section{An approximation for the optical depth}
\label{sec:appendixb}

In order to estimate the optical depth of the HI gas in our simulations, we repeat one radiative transfer calculation for MC1-HD, however, now setting the background temperature to 0~K. Hence, the observed emission purely stems from the HI gas in the zoom-in region. 

As noted in Eq.~\ref{eq:NH}, but now written down for a single velocity channel, the HI column density is given by
\begin{equation}
 \rmn{d}N_\rmn{HI} = 1.8224 \times 10^{18} \text{cm$^{-2}$} \, \frac{T_\rmn{s}}{\text{1 K}} \tau(v) \frac{\rmn{d}v} { \textrm{1 km s$^{-1}$} } \, .
 \label{eq:dNH}
\end{equation}
Next, we consider the expression
\begin{equation}
T_\rmn{rad} = \frac{h \nu_\rmn{HI}}{k_\rmn{B}} \left( f(T_\rmn{s}) - f(T_\rmn{bg}) \right) \left(1 - e^{-\tau} \right) \, ,
\label{eq:Tradid}
\end{equation}
for the observed brightness temperature $T_\rmn{rad}$. Here, $T_\rmn{bg}$ is the background temperature, $h$ is the Planck constant, $k_\rmn{B}$ the Boltzmann constant, $\nu_\rmn{HI}$ = 1420 MHz the frequency of the HI 21-cm line, and
\begin{equation}
 f(T) = \frac{1}{\rmn{exp}\left( \frac{h \nu_\rmn{HI}}{k_\rmn{B} T} \right) - 1} \, .
\end{equation}
Considering that \mbox{$h \nu_\rmn{HI}/k_\rmn{B} = 0.068$ K $\ll T_\rmn{s}$} and \mbox{$f(T_\rmn{bg}) \ll f(T_\rmn{s})$}, and inserting Eq.~\ref{eq:Tradid} in Eq.~\ref{eq:dNH} yields
\begin{equation}
 \rmn{d}N_\rmn{HI} \simeq 1.8224 \times 10^{18} \text{cm$^{-2}$} \, \frac{\tau}{1-e^{-\tau}} \frac{T_\rmn{rad}}{\text{1 K}}  \frac{\rmn{d}v} { \textrm{1 km s$^{-1}$} } \, .
\end{equation}
We now integrate over all velocity channels using a definition of a $T_\rmn{rad}$-weighted, channel-averaged approximation of the optical depth
\begin{equation}
 N_\rmn{HI} = 1.8224 \times 10^{18} \text{cm$^{-2}$} \, \langle \tau \rangle \int \frac{T_\rmn{rad}}{\text{1 K}}  \frac{\rmn{d}v} { \textrm{1 km s$^{-1}$} } \, .
\end{equation}
Here, we have defined
\begin{equation}
 \langle \tau \rangle  = \frac{\int \frac{\tau}{1-e^{-\tau}} T_\rmn{rad} \rmn{d}v}{\int T_\rmn{rad} \rmn{d}v} \, .
\end{equation}
The interpretation of $\langle \tau \rangle$ as an approximation for a $T_\rmn{rad}$-weighted, channel-averaged optical depth can be understood, when considering the fact that \mbox{$\frac{\tau}{1-e^{-\tau}} \rightarrow \tau$} with \mbox{$\tau \rightarrow \infty$}. For \mbox{$\tau$ = 1}, the expression $\frac{\tau}{1-e^{-\tau}}$ is only $\sim$50\% larger than $\tau$, for \mbox{$\tau$ = 2} only $\sim$15\%. For optically thin regions ($\tau < 1$), the approximation is not applicable. However, as in Section~\ref{sec:optdepth} we are mainly interested in high optical depth regions, we consider our definition of $\langle \tau \rangle$ as a reasonable approximation for the typical optical depth of HI in our simulations.

Next, using the integrated intensity from the radiative transfer calculations without any background radiation field and the real HI column density from the simulation data, $N_\rmn{HI, real}$, we can now calculate $\langle \tau \rangle$ via
\begin{equation}
 \langle \tau \rangle = \frac{N_\rmn{HI, real}}{1.8224 \times 10^{18} \text{cm$^{-2}$} \, \int \frac{T_\rmn{rad}}{\text{1 K}}  \frac{\rmn{d}v} { \textrm{1 km s$^{-1}$} } } \, ,
\end{equation}
where the denominator is describing the HI column density obtained from HI emission under the assumption of optically thin emission.

%


\bsp	
\label{lastpage}
\end{document}